\documentclass[10pt]{iopart}

\usepackage{iopams}

\expandafter\let\csname equation*\endcsname\relax
\expandafter\let\csname endequation*\endcsname\relax
\usepackage{amsmath}
\usepackage{graphicx}

\usepackage{xcolor}
\usepackage{soul}

\begin{document}

\title[]{Quasi-harmonic temperature dependent elastic constants: 
applications to silicon, aluminum, and silver}


\author{Cristiano Malica and Andrea Dal Corso}

\address{International School for Advanced Studies (SISSA), \\
Via Bonomea 265, 34136 Trieste (Italy).}
\ead{cmalica@sissa.it}
\vspace{10pt}
\begin{indented}
\item[]December 2019
\end{indented}

\begin{abstract}
We present ab-initio calculations of the quasi-harmonic temperature dependent
elastic constants. The isothermal elastic constants are calculated at each
temperature as second derivatives of the Helmholtz free energy with respect
to strain and corrected for finite pressure effects. This calculation is
repeated for a grid of geometries and the results interpolated at the minimum
of the Helmholtz free energy. The results are compared with the quasi-static
elastic constants. Thermodynamic relationships are used to derive the
adiabatic elastic constants that are compared with the experimental measurements.
These approaches are implemented for cubic solids in the \texttt{thermo\_pw} code
and are validated by applications to silicon, aluminum, and silver.
\end{abstract}

%
%
%
%
\ioptwocol

\section{Introduction}

Elastic constants (ECs) characterize the mechanical and thermodynamic behaviors of
materials. They determine the speed of sound, the crystal stability, and allow the
calculation of properties such as the thermal expansion (TE) or the thermal stresses.
They have also practical applications as, for instance, the prediction of seismic
properties of materials, a basic information to probe the interior of planets in
geophysics~\cite{Gillan_2006}.

First-principles calculations of the ECs~\cite{Nielsen_1985} help to interpret experiments and
might be useful complements at extreme conditions of temperature and
pressure that are difficult or impossible to realize in laboratory.
Density Functional Theory (DFT) has been employed for several decades to estimate the
ECs of materials and usually provides values that are within $\approx$ 10\% of the experiments
\cite{Wang_2009,Zhao_2007,Golesorkhtabar_2013,Perger_2010,Jamal_2014,Jamal_2018}.

Temperature might have a not negligible effect on ECs, nevertheless DFT
calculations are usually limited to zero temperature ($T=0$ K) since
a significant computational effort is necessary to evaluate the
ECs including both temperature and quantum effects.
In the literature there are two main approaches based on lattice dynamics and the
computation of the phonon dispersions of solids. In the quasi-static approximation (QSA) temperature dependent ECs
(TDECs) are computed assuming that temperature produces only a TE. At each
temperature the ECs are calculated at the geometry that minimizes the Helmholtz free energy
but the ECs themselves are calculated as second derivatives of the DFT total energy with
respect to strain or from the derivative of the stress with respect to strain (at $T=0$ K).
For instance, Y. Wang {\it et al.} computed the TDECs of seven
cubic metals~\cite{Wang_2010} and S-L. Shang {\it et al.} ~\cite{Shang_2010} computed TDECs of $\alpha$-
and $\theta$-Al$_2$O$_3$ within the QSA. K. K\'adas {\it et al.}~\cite{kadas_2007}
used the same approximation for the TDECs of $\alpha$-beryllium deriving the
temperature dependence of the volume from the Debye model.
Within the second approach, which is based on the quasi-harmonic approximation (QHA),
the ECs are calculated from second derivatives of the Helmholtz free energy
with respect to strain
and can be calculated at the temperature dependent geometry. This
approximation has been applied for instance to the TDECs of
MgO~\cite{mgo, karki}, hexagonal close-packed (h.c.p.) beryllium and cubic and h.c.p.
diamond~\cite{Shao_2012},
$\alpha$-iron~\cite{dragoni} and Fe$_3$Ga alloys ~\cite{yan}.
Recently, M. Destefanis {\it et al.}~\cite{erba}
computed TDECs for the Forsterite mineral using the QHA.

In this paper we present our implementation of TDECs in the \texttt{thermo\_pw}
code~\cite{tpw}, a driver of \texttt{Quantum ESPRESSO}~\cite{qe1, qe2} routines
for the calculation of the thermodynamic properties of solids.
This code has been used for the $T=0$ K elastic constants of beryllium ~\cite{Dal_Corso_2016},
the thermodynamic properties of h.c.p. metals rhenium, technetium,~\cite{pal} ruthenium, and osmium~\cite{pal2} 
and recently to account for anharmonic contributions to the mean square atomic displacements within
the QHA~\cite{bf}
\footnote{ 
Further tests are reported in the supplementary material: we 
show the computed $T=0$ K ECs of indium, TiO$_2$ rutile, 
$\alpha$-Al$_2$O$_3$.}.
We implemented TDECs both within the QSA and the QHA. The latter can be calculated at
a single reference geometry or for a set of reference geometries and further
interpolated at the temperature dependent geometry.

We validate our implementation by studying the TDECs of three elemental solids:
silicon with the diamond structure and the face-centered cubic metals
aluminum and silver
\footnote{Moreover in the supplementary material we check our method by computing the TDECs of MgO within the QHA 
and by comparing our results with those of Refs.~\cite{mgo,karki}.}.
Derivatives of the Helmholtz free energy give the isothermal ECs,
while usually experiments measure the adiabatic ECs. We calculate the
latter through thermodynamic relationships and find good agreement
with the available experiments.
The low temperature ECs ($T=77$ K for silicon, $T=0$ K for aluminum and silver) are
within $\approx 10 \%$ from experimental values for both QSA and QHA.
Increasing temperature, the QHA gives almost the same percentage change as the
experiment while the QSA gives a somewhat smaller softening.

Finally we tested the accuracy of our TDECs by comparing the TE computed using
mode-Gr\"{u}neisen parameters that depend on the TDECs through the bulk
modulus with the
TE obtained by differentiation of the temperature dependent lattice
constant which we take as a reference.
We find that the two methods are in good agreement when using the QHA
TDECs, while some differences are introduced by the QSA.
In the literature the TE is usually evaluated from mode-Gr\"{u}neisen parameters,
using temperature independent ECs (for instance the $T=0$ K ECs) since this approach
is numerically more efficient than the differentiation of the crystal parameters
especially in anisotropic solids.
We computed the TE also within this approximation finding reasonable agreement
with the reference but discrepancies larger than
those obtained by the QHA or the QSA TDECs.

\section{Theory}
We consider a crystal deformed with a symmetric strain $\epsilon_{ij}$, where
$i$ and $j$ are Cartesian coordinate indices. To maintain the solid
in a strained configuration, forces have to be applied in order to balance those
exerted by the solid. These forces per unit area give the
stress tensor which, for small strains, is proportional to the strain:
\begin{equation}\label{sigma}
\sigma_{ij} = \sum_{kl} C_{ijkl} \epsilon_{kl}, 
\end{equation}
where $C_{ijkl}$ are the components of the ECs tensor. Equivalently:
\begin{equation}\label{ss}
C_{ijkl} = \left(\frac{\partial \sigma_{ij}}{\partial \epsilon_{kl}}\right)_{\epsilon=0}.
\end{equation}
At $T=0$ K the ECs can be written also as the second derivatives
of the energy with respect to strain:
\begin{equation}\label{dU}
\tilde C_{ijkl} = \frac{1}{\Omega} \left(\frac{\partial^2 U}{\partial \epsilon_{ij} 
\partial \epsilon_{kl}} \right)_{\epsilon=0},
\end{equation}
where $\Omega$ is the volume of the reference system and $U$ is the DFT total
energy. If the reference geometry minimizes the energy (the equilibrium condition)
$\tilde C_{ijkl}=C_{ijkl}$, while for a reference geometry in which a pressure $p$
is acting on the solid (in this paper we do not consider the possibility of having
a generic stress on the solid) $C_{ijkl}$
and $\tilde C_{ijkl}$ differ and we have~\cite{Barron_1965}:
\begin{equation}\label{dUp}
C_{ijkl} = \tilde C_{ijkl} + \frac{1}{2} p \left(2 \delta_{ij} \delta_{kl} - \delta_{il} \delta_{jk} 
- \delta_{ik} \delta_{jl} \right).
\end{equation}
Eqs.~\ref{ss} and~\ref{dUp} provide two equivalent ways to compute the
stress-strain ECs $C_{ijkl}$ at $T=0$ K.

In cubic solids and in Voigt notation (see for instance chapter
VIII of~\cite{nyebook}) the ECs tensor has the following form:
\begin{equation} \label{ec_matrix}
\left( \begin{array}{cccccc}
\tilde C_{11} & \tilde C_{12} &  \tilde C_{12} & . & . & .
\\
\tilde C_{12} & \tilde C_{11} & \tilde C_{12} & . & . & .
\\
\tilde C_{12} & \tilde C_{12} & \tilde C_{11} & . & . & .
\\
. & . & . & \tilde C_{44} & . & . 
\\
. & . & . & . & \tilde C_{44} & .
\\
. & . & . & . & . & \tilde C_{44} 
\end{array}
\right),
\end{equation}
with three independent ECs, $\tilde C_{11}$, $\tilde C_{12}$, and $\tilde C_{44}$ (a dot
indicates a zero entry).

From Eq.~\ref{dU} the total energy (at $T=0$ K) contains a term quadratic in
the strain:
\begin{equation} \label{u}
U = \frac{\Omega}{2} \sum_{ij} \epsilon_i \tilde C_{ij} \epsilon_j.
\end{equation}

In order to derive the three independent ECs we use the following strains:

\begin{equation} \label{strains}
\begin{aligned}
\epsilon_A & =\left( \begin{array}{ccc}
\epsilon_{1} & 0 & 0 
\\
0 & \epsilon_{1} &  0 
\\
0 & 0 & \epsilon_{1}
\end{array}
\right),\ \ 
\epsilon_{E}=\left( \begin{array}{ccc}
0 & 0 & 0 
\\
0 & 0 &  0
\\
0 & 0 & \epsilon_{3}
\end{array}
\right), \\
\epsilon_{F} & =\left( \begin{array}{ccc}
0 & \epsilon_{4} & \epsilon_{4} 
\\
\epsilon_{4} & 0 & \epsilon_{4}
\\
\epsilon_{4} & \epsilon_{4} & 0
\end{array}
\right).
\end{aligned}
\end{equation}

The strain $\epsilon_A$ does not change the shape of the cubic cell, 
while $\epsilon_E$ transforms it into a tetragonal cell and $\epsilon_F$
into a rhombohedral cell.

Since the $\tilde C_{ij}$ tensor has the form as in Eq.~\ref{ec_matrix},
applying Eq.~\ref{u} we obtain for each strain the following relationships:
\begin{equation} \label{uA}
\begin{aligned}
U_A & = \frac{3 \Omega}{2} \left(\tilde C_{11}+ 2 \tilde C_{12}\right) {\epsilon_1}^2,
\\
U_E & = \frac{\Omega}{2} \tilde C_{11} {\epsilon_3}^2,
\\
U_F & = \frac{3 \Omega}{2} \tilde C_{44} {\epsilon_4}^2.
\end{aligned}
\end{equation}
We compute the ECs $\tilde C_{ij}$ by fitting these equations with polynomials and taking
the analytic second derivatives.

In order to introduce zero-point quantum effects and
temperature, we can use the same formulation by replacing in
Eq.~\ref{dU} the total energy $U$ with the Helmholtz free energy $F$:
\begin{equation}\label{dUT}
\tilde C_{ijkl}^T = \frac{1}{\Omega} \left(\frac{\partial^2 F}{\partial \epsilon_{ij} 
\partial \epsilon_{kl}} \right)_{\epsilon=0}.
\end{equation}
The Helmholtz free energy is the sum of the total energy $U$ and the vibrational 
free energy (neglecting the electronic contribution): $F=U+F_{vib}$. 
The latter is given by:
\begin{eqnarray} \label{equ4}
F_{vib}(\mathbf \epsilon, T) & = \frac{1}{2N} \sum_{\mathbf q \eta} \hbar \omega_{\eta} \left(\mathbf q,
\mathbf \epsilon \right) \nonumber\\
& + \frac{k_B T}{N} \sum_{\mathbf q \eta} \ln \left[1 - \exp \left(-\frac{\hbar \omega_{\eta}
(\mathbf q, \mathbf \epsilon)}{k_B T}\right) \right],
\end{eqnarray}
where the sums are over the phonon modes identified by 
wave-vectors ${\bf q}$ in the first Brillouin zone and mode indices $\eta$. 
$\omega_{\eta}\left(\mathbf q, \epsilon \right)$ indicates the phonon frequencies,
$\Omega$ the volume of the reference unit cell,
$k_B$ the Boltzmann constant, $\hbar$ the Planck constant divided by $2 \pi$, 
$T$ the absolute temperature and $N$ the number of unit cells in the solid. 
The first term on the right-hand side is the zero-point energy while the second 
is the vibrational contribution at finite temperature.

Within the QHA the phonon frequencies depend on the applied strain $\epsilon$.
In the QHA calculation at fixed reference geometry the phonon dispersions 
and free energies are computed at different strained configurations. 
For each temperature the free energy-strain function is fitted with polynomials 
whose second derivatives at zero strain give the $\tilde C_{ijkl}^T$ ECs of Eq.~\ref{dUT}. 
In general the reference configuration is not at the minimum of the Helmholtz free 
energy so there is a pressure $p=-{d F \over d \Omega}$ and we correct the
$\tilde C_{ijkl}^T$ with the generalization of Eq.~\ref{dUp} 
at finite temperature in order to obtain the $C_{ijkl}^T$.
Finally we call QHA TDECs those obtained by performing QHA calculations on
a few fixed reference geometries and interpolating the $C_{ijkl}^T$ at each temperature $T$ 
at the geometry that minimizes the Helmholtz free energy.
In other works (see for instance~\cite{dragoni}) the QHA ECs are evaluated with
a slightly different procedure. First for each type of strain
the Helmholtz free energy is fitted by a polynomial in a grid of the lattice constants
and strain values and then the ECs are calculated as second derivatives 
of the Helmholtz free-energy with respect to the strain at the temperature 
dependent lattice constant. 
We have verified that the results obtained by this more conventional method 
are very similar to those obtained by our procedure and both approaches 
are available in \texttt{thermo\_pw}.

QSA TDECs are derived in a similar way using the second derivatives of $U$ instead 
of the derivatives of $F$, in this way avoiding to compute the phonon dispersions 
for all strained configurations. The phonon dispersions are instead computed at 
the reference geometries in order to obtain the temperature dependent crystal 
parameters.

These procedures give for each temperature the isothermal ECs $C_{ijkl}^T$. However, 
many experimental setups using for instance ultrasonic pulse techniques, measure the 
adiabatic ECs $C_{ijkl}^S$. The latter can be readily obtained using the 
thermal stresses $b_{ij}$  and the isochoric heat capacity $C_{V}$~\cite{wallace}:
\begin{equation} \label{adiab}
C_{ijkl}^S = C_{ijkl}^T + \frac{T \Omega b_{ij} b_{kl}}{C_{V}}.
\end{equation}
$b_{ij}$ is obtained from:
\begin{equation} \label{equ6}
b_{ij} = - \sum_{kl} C_{ijkl}^T \alpha_{kl},
\end{equation}
where $\alpha_{kl}$ is the TE tensor that, in the cubic case is 
isotropic: $\alpha_{kl}=\alpha \delta_{kl}$. We have:
\begin{equation} \label{alfa}
\alpha = \frac{1}{a(T)} \frac{d a(T)}{d T},
\end{equation}
where $a(T)$ is the temperature dependent cubic lattice constant. 
The calculation of $a(T)$ as well as the calculation of $C_V$
are explained in Ref.~\cite{pal} and summarized in the supplementary material.

As a further check of the consistence of our ECs we recalculate the TE using
the mode-Gr\"uneisen parameters. This formula requires the isothermal
bulk modulus that can be obtained either from the QSA or from the QHA TDECs.
The resulting TE can be compared with the TE given by Eq.~\ref{alfa}.
In cubic solids the TE written in terms of the Gr\"uneisen parameters is given by
\begin{equation} \label{alfagrun}
\alpha = - \frac{1}{3 B^T} \sum_{\mathbf q \eta} \gamma_{\eta}(\mathbf q) c_{\eta}(\mathbf q).
\end{equation}
The $\gamma_{\eta}(\mathbf q)$ are the mode-Gr\"{u}neisen parameters:
\begin{equation} \label{gamma}
\gamma_{\eta}(\mathbf q) = - \frac{\Omega}{\omega_{\eta}(\mathbf q)} \frac{\partial \omega_{\eta}
(\mathbf q)}{\partial \Omega},
\end{equation}
and $c_{\eta}(\mathbf q)$ are the mode contributions to the isochoric specific 
heat:
\begin{equation} \label{c}
c_{\eta}(\mathbf q) = \frac{\hbar \omega_{\eta}(\mathbf q)}{\Omega}
\frac{\partial}{\partial T} \left({\e^{\frac{\hbar \omega_{\eta}(\mathbf q)}{k_B T}}-1}\right)^{-1}.
\end{equation}
In order to apply Eq.~\ref{alfagrun} the phonon frequencies 
calculated at the selected
geometries are interpolated with a polynomial as a function of the lattice constant.
Then, for each temperature T, we evaluate the interpolating polynomial (to
obtain the frequencies) and its first derivative (proportional to the
mode-Gr\"{u}neisen parameters) at the $a(T)$. The $c_{\eta}(\mathbf q)$ are
evaluated from the interpolated frequencies at that temperature.
$B^T$ is the isothermal bulk modulus which for cubic solids can be written 
in terms of the isothermal ECs as:
\begin{equation} \label{bmod}
B^T = \frac{1}{3} \left(C_{11}^T + 2 C_{12}^T \right).
\end{equation}
This bulk modulus can be compared also with the bulk modulus
derived at each temperature from the fit of the free-energy as a function of
the volume with the Murnaghan equation.
The calculation of the TE using Eq.~\ref{alfagrun} and TDECs is not very efficient
since it requires first the evaluation of the temperature dependent crystal
parameters (and hence implicitly of the TE itself). This problem is usually
solved in the literature neglecting the variation of the ECs with temperature.
We verified also this approximation as discussed below.

Finally the isothermal bulk modulus can be converted into the adiabatic one with the
following formula:
\begin{equation} \label{bmod_adab}
K^T - K^S = \frac{T \Omega \beta^2}{C_P}
\end{equation}
where $K^T=B^{T-1}$ is the isothermal compressibility,
$K^S=B^{S-1}$ is the isobaric compressibility,
$\beta=3 \alpha$ is the volume
thermal expansion and $C_P$ is the isobaric heat capacity computed as in Ref.
\cite{pal}.

A flow-chart and further details on the computation of the 
TDECs with the
\texttt{thermo\_pw} code are reported in the supplementary material.

\begin{table*} \centering
\caption{Theoretical lattice parameters (in \AA) at different temperatures
compared with experiment~\cite{wyckoff} at room temperature. ZPE indicates the
zero point energy.
}
\begin{tabular}{lcccc}
\hline
& \multicolumn{1}{c}{T=0 K} \ \ & \multicolumn{1}{c}{T=0 K+ZPE} \ \ & \multicolumn{1}{c}{T=300 K} \ \ & \multicolumn{1}{c}{Expt.} \\
\hline
\hline
Silicon  (LDA)     \  & $5.40$ & $5.41$ & $5.41$ & $5.4307$ \\
Aluminum (PBEsol)  \  & $4.01$ & $4.03$ & $4.04$ & $4.04958$ \\
Silver   (PBEsol)  \  & $4.06$ & $4.07$ & $4.08$ & $4.0862$ \\
\hline
\hline
\end{tabular}
\label{table1}
\end{table*}

\begin{table*} \centering
\caption{ECs at $T=0$ K compared with the results available in the literature. The exchange and correlation functionals are indicated
in the first column. In addition to PBEsol used by us, the GGA functionals PBE~\cite{pbe} and PW~\cite{pw} were used in the references.
The equilibrium lattice constants ($a_0$) is in \AA\ while the ECs are in kbar.
}
\begin{tabular}{lcccc}
\hline
& \multicolumn{1}{c}{$a_0$} \ \ & \multicolumn{1}{c}{$C_{11}$} \ \ & \multicolumn{1}{c}{$C_{12}$} \ \ & \multicolumn{1}{c}{$C_{44}$} \\
\hline
\hline
Silicon \\
\hline
LDA$^a$ & 5.40 & 1618 & 640 & 761 \\
LDA$^b$ & 5.41 & 1580 & 639 & 746 \\
LDA$^c$ & 5.40 & 1590 & 610 & 850 \\
LDA$^d$ & 5.38 & 1621 & 635 & 773 \\
LDA$^e$ & 5.40 & 1610 & 650 & 760 \\
PBE$^e$ & 5.47 & 1530 & 570 & 740 \\
PBEsol$^e$ & 5.44 & 1560 & 620 & 740 \\
Expt.$^l$ & 5.43 & 1675 & 650 & 801 \\
\hline
\hline
Aluminum \\
\hline
PBEsol$^a$ & 4.01 & 1192 & 643 & 365 \\
PBEsol$^b$ & 4.03 & 1146 & 632 & 353 \\
LDA$^g$ & 3.97 & 1222 & 608 & 374 \\
LDA$^h$ & 4.04 & 1104 & 545 & 313 \\
PBE$^f$ & 4.06 & 1093 & 575 & 301 \\
PW$^i$ & 4.05 & 1010 & 610 & 254 \\
Expt.$^m$ & 4.05 & 1143 & 619 & 316 \\
\hline
\hline
Silver \\
\hline
PBEsol$^a$ & 4.06 & 1450 & 1067 & 540 \\
PBEsol$^b$ & 4.07 & 1425 & 1054 & 531 \\
GGA$^h$ & 4.02 & 1612 & 1191 & 581 \\
PW$^i$ & 4.16 & 1159 & 851 & 421 \\
Expt.$^n$ & 4.09 & 1315 & 973 & 511 \\
\hline
\end{tabular}
\label{table2}
\\ $^a$ This work at $T=0$ K,
$^b$ This work at $T=0$ K + ZPE,
$^c$ Reference~\cite{Nielsen_1985},
$^d$ Reference~\cite{Zhao_2007},
$^e$ Reference~\cite{rasander_2015}
\\ $^f$ Reference~\cite{Golesorkhtabar_2013},
$^g$ Reference~\cite{weixue_1998},
$^h$ Reference~\cite{Wang_2009},
$^i$ Reference~\cite{Shang_2010},
\\ $^l$ $a_0$ from Table 1 at $T=300$ K. ECs at $T=77$ K~\cite{si_ec_exp}
\\ $^m$ $a_0$ from Table 1 at $T=300$ K. ECs extrapolated at $T=0$ K~\cite{kamm}
\\ $^n$ $a_0$ from Table 1 at $T=300$ K. ECs extrapolated at $T=0$ K~\cite{Neighbours_1958}
\end{table*}

\section{Method}

The calculations presented in this work were carried out using DFT as implemented
in the Quantum Espresso package ~\cite{qe1, qe2}. The exchange and correlation functional is approximated
by the local density approximation (LDA) for silicon~\cite{lda} and the generalized gradient approximation of
Perdew-Burke-Ernzerhoff modified for densely packed solids (PBEsol)~\cite{pbesol} for aluminum and silver.
We took the exchange and correlation functionals that, on the basis of previous theoretical calculations, give a better
agreement with the experiment for the TE and/or phonon dispersion curves.
 
We employed the projector augmented wave (PAW) method and a plane waves basis set with
pseudopotentials~\cite{paw} from $pslibrary$ ~\cite{psl}.
For silicon we used the pseudopotential \texttt{Si.pz-nl-} \texttt{kjpaw\_psl.1.0.0.UPF}, the cutoff for the wave functions
was $60$ Ry, the one for the charge density $640$ Ry, and the \textbf{k}-points mesh was $16 \times 16 \times 16$.
For aluminum we used the pseudopotential \texttt{Al.pbesol-n-kjpaw\_psl.1.0.0.UPF}, the cutoff for the wave functions
was $30$ Ry, the one for the charge density $120$ Ry, and the \textbf{k}-points mesh was $48 \times 48 \times 48$.
For silver we used the pseudopotential \texttt{Ag.pbesol-n-kjpaw\_psl.1.0.0.UPF}, the cutoff for the wave functions
was $70$ Ry, the one for the charge density $300$ Ry, and the \textbf{k}-points mesh was $64 \times 64 \times 64$.
In the case of aluminum and silver, the presence of the Fermi surface has been dealt with by a smearing approach
~\cite{mp} with a smearing parameter $\sigma = 0.02$ Ry. Density functional perturbation theory (DFPT)~\cite{rmp, dfptPAW} 
was used to calculate the dynamical matrices on a $8 \times 8 \times 8$ \textbf{q}-points grid
for silicon while for the metals we employed a $4 \times 4 \times 4$ \textbf{q}-points grid.
These dynamical matrices have been Fourier interpolated on a $200 \times 200 \times 200$ \textbf{q}-points mesh
to evaluate the free-energy.

The calculation of the TDECs can be done automatically by the \texttt{thermo\_pw} program. The user chooses the relevant parameters:
the number of strained configurations, the interval of strain values between two geometries, the degree of the
fitting polynomials used in the calculation (further details for the calculations are reported
in the \texttt{thermo\_pw} user's guide and in the file $thermo.pdf$, released with the package).
For our three materials the grid of the reference geometries were centered at the $T = 0$ K lattice constants reported
in Table~\ref{table1}.
We used $9$ reference geometries for silicon with lattice constants separated from
each other by $\Delta a= 0.05$ a.u.,
$7$ for aluminum separated by $\Delta a = 0.07$ a.u., and
$7$ for silver separated by $\Delta a = 0.085$ a.u..
The same grid is used for the calculation of the $a(T)$ and the TE.
For all three materials we used 6 strained configurations for each type of strain
with a strain interval $\delta \epsilon=0.005$. In total we computed the phonon
dispersions for $162$ geometries for silicon and $126$ geometries for aluminum and
silver.
In order to fit the energy (for the QSA) or the
free-energy (for the QHA) as a function of strain we use a polynomial of degree two.
To fit the ECs computed at the various reference configurations at the temperature dependent geometry we use a polynomial of degree four.

\section{Applications}
In Table~\ref{table1} we report the theoretical lattice constants for silicon, aluminum, and silver:
the equilibrium values at $T=0$ K with and without zero point energy and the values
at $T=300$ K. The latter is in good agreement with the room temperature
experimental value~\cite{wyckoff}:
for all three elements, the experimental values being slightly larger (with differences
of $0.03$ \AA\ in silicon, $0.01$ \AA\ in aluminum, and even less in silver).

\begin{figure*}
\centering
\includegraphics[width=0.49\linewidth]{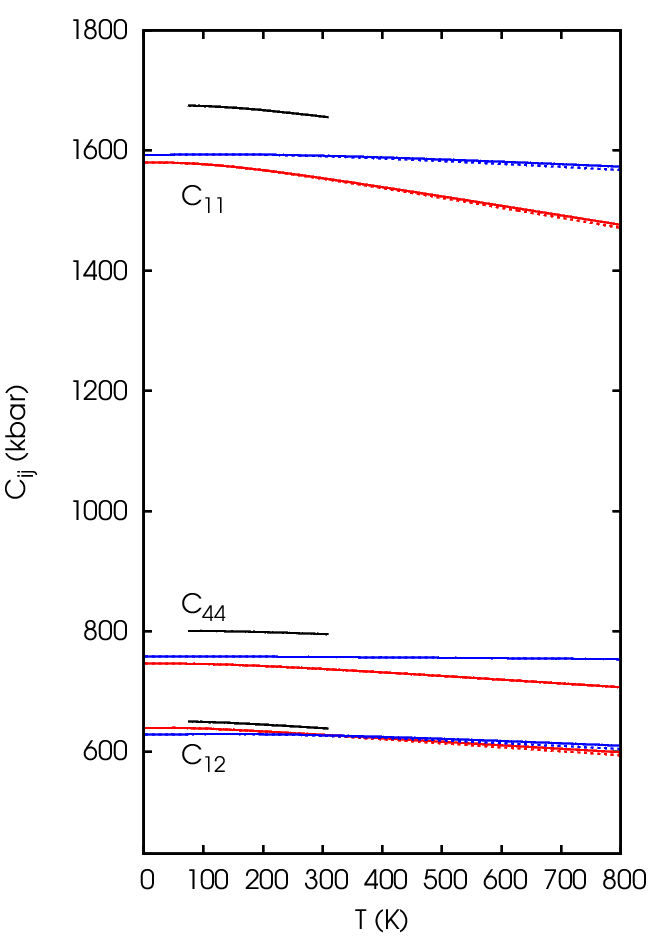}
\caption{Temperature dependent elastic constants of silicon. QHA (red curves) is compared with
        QSA (blue curves). The isothermal (dashed lines) and adiabatic (continuous lines) elastic
        constants are reported. Experimental data are taken from McSkimin~\cite{si_ec_exp} (black lines).}
\label{fig:si_ec}
\end{figure*}

In Table~\ref{table2} we compare our ECs computed at $T=0$ K with those obtained considering the zero point
energy given by QHA. We also report the comparison with experiment and other works.
In Figure~\ref{fig:si_ec} the TDECs of silicon computed by means of the QSA (blue curves) and the
QHA (red curves) are shown. The dashed lines indicate isothermal ECs, the continuous lines the
adiabatic ECs obtained by Eq.~\ref{adiab}.
The experimental points (black line) are adiabatic ECs obtained from ultrasonic experiments
~\cite{si_ec_exp}.

At the temperature corresponding to the experimental points the difference between isothermal
and adiabatic ECs $C_{11}$ and $C_{12}$ is quite small and it remains small also at higher temperatures.
For $C_{44}$ there is no difference as expected for cubic solids with
a diagonal TE tensor.
In general the theoretical values are slightly below the experimental data (see also Table~\ref{table2}):
in particular these differences are about $6$ \% for $C_{11}$, $2$ \%
for $C_{12}$, and $7$ \% for $C_{44}$ at the lowest experimental temperature $T=77$ K.
Our $T=0$ K EC agrees with almost all LDA ECs reported in Table~\ref{table2} with typical differences of about 2 \%. Only Ref.~\cite{Nielsen_1985} has larger discrepancies.
The PBE and PBEsol values are lower than LDA and hence more distant from experiment.
The experimental ECs decreases
by $\approx 1.1 \%$ for $C_{11}$, $\approx 1.8 \%$ for $C_{12}$, and
$\approx 0.7 \%$ for
$C_{44}$ from $77$ K to $300$ K. The theoretical softening in the
same temperature range are $\approx 1.6 \%$ (QHA) and $\approx 0.1 \%$
(QSA) for $C_{11}$, $\approx 1.9 \%$ (QHA) and $\approx 0.3 \%$ (QSA) for $C_{12}$,
$\approx 1.2 \%$ (QHA) and $\approx 0.1 \%$ (QSA) for $C_{44}$.
The theoretical QHA reproduces better the experimental trend than the QSA.

\begin{figure*}
\centering
\includegraphics[width=.49\linewidth]{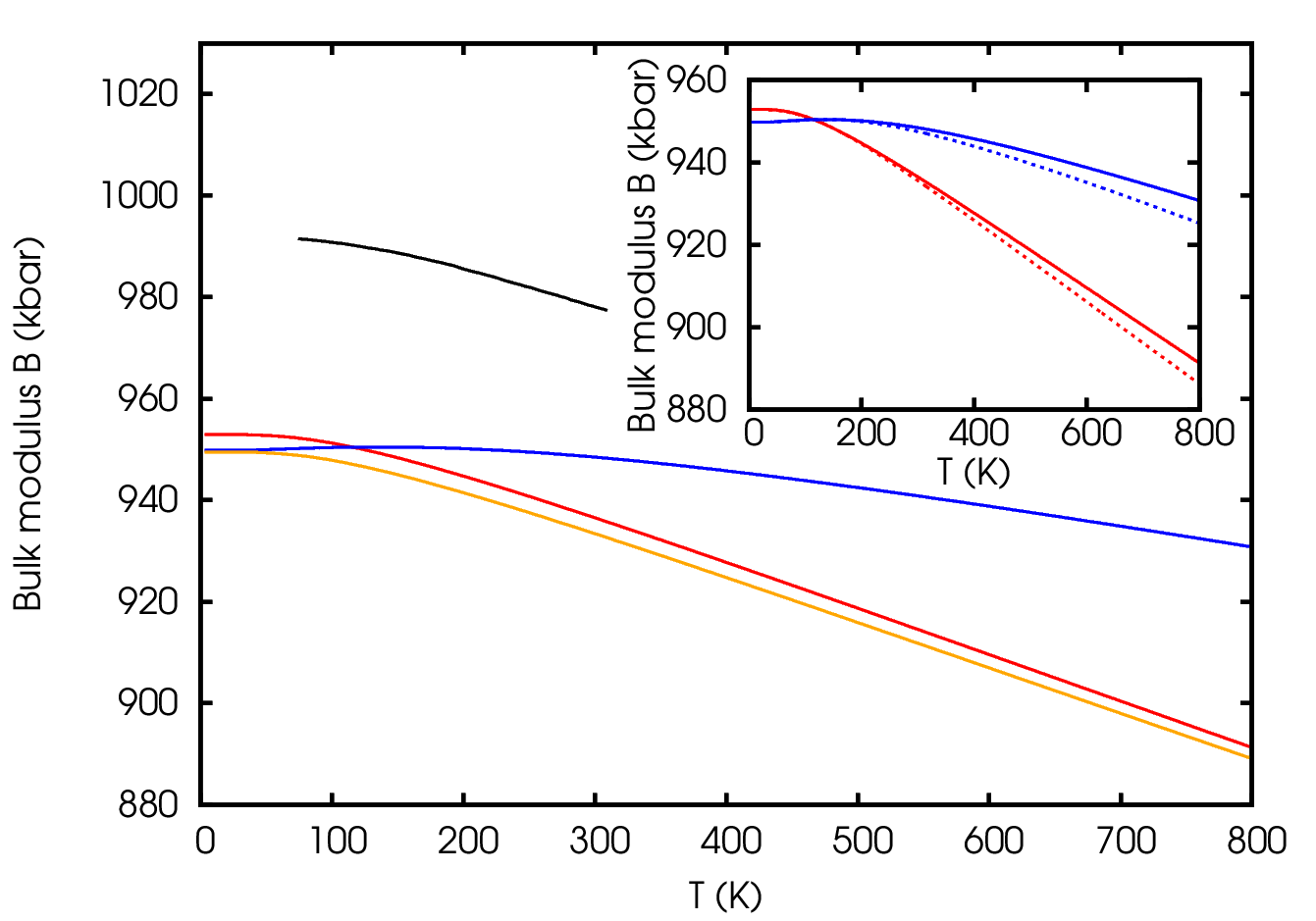}
\includegraphics[width=.49\linewidth]{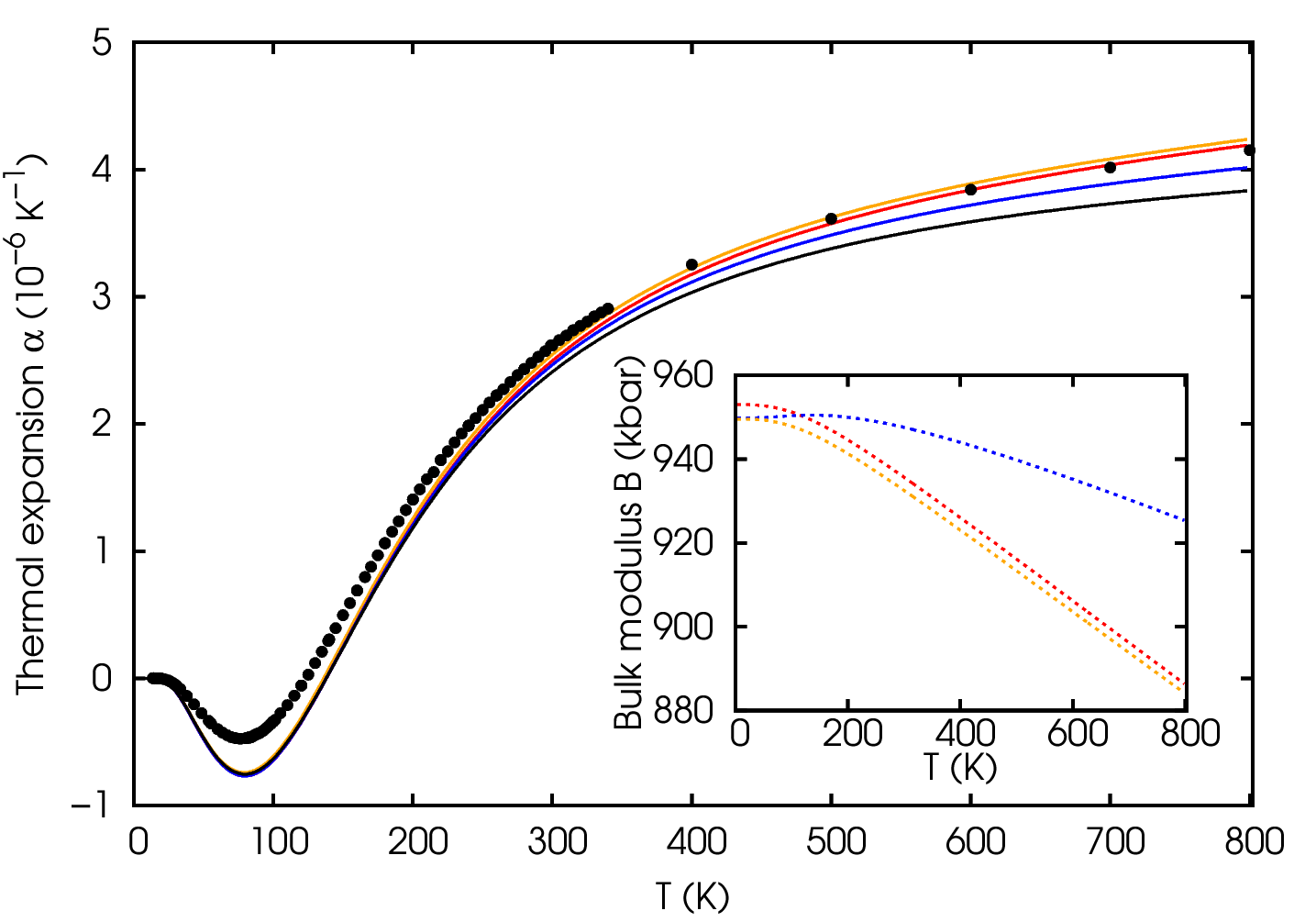}
\caption{Left. Temperature dependent bulk modulus of silicon. QHA (red curves), QSA (blue curves) and experimental data (black line)~\cite{si_ec_exp}.
        The orange curve is the bulk modulus obtained from the Murnaghan equation. In the inset a comparison between the
        adiabatic (continuous lines) and the isothermal (dashed lines) bulk moduli.
        Right. Thermal expansion of silicon: computed as in Eq.~\ref{alfa} (orange curve). The other curves
        are computed by using Eq.~\ref{alfagrun} with a bulk modulus derived from ECs (Eq.~\ref{bmod}):
        TDECs computed via QHA (red curve) or QSA (blue curve) and $T=0$ ECs
        (black curve). The experimental points are taken from~\cite{si_te_1} and~\cite{book3}. In the inset the isothermal bulk moduli are reported:
        QHA (red dashed line), QSA (blue dashed line) and obtained from Murnaghan equation (orange dashed line).}
\label{fig:si_te}
\end{figure*}

Although in the experimental temperature range the decrease of the ECs is small,
in the whole temperature range $0$ K - $800$ K considered in the plot, the QHA
ECs have a not negligible variation:
$C_{11}$ and $C_{12}$ decrease of about 7 \% and $C_{44}$ of 5 \%.
In Figure~\ref{fig:si_te} we show the bulk modulus obtained from these TDECs.
The comparison between theoretical and experimental adiabatic bulk moduli
obtained from the experimental ECs using Eq.~\ref{bmod} (left) reflects the
behavior of the ECs: the temperature dependence is more in agreement with QHA (red curve) than QSA (blue curve).
Moreover, the QHA is much more in agreement with the bulk modulus obtained from the
Murnaghan equation (orange curve) and differs from the experiment (black line) by an almost constant amount.

Finally we present the TE calculated by the
Gr\"{u}neisen's formula (Eq.\ref{alfagrun}).
In order to gauge our TDECs we take as reference the TE coefficient computed
via Eq.~\ref{alfa}.
This curve is reported in the right side of Figure~\ref{fig:si_te} (orange).
The difference of the TEs with the reference is quantified by computing the area percentage error (APE) that is the
percentage difference of the areas under the TE curves.
The TE obtained with the isothermal bulk modulus derived from the Murnaghan equation (orange dashed line in the inset)
differs by less than 0.04\%.
The TE derived from QHA almost overlaps with the previous one (it is slightly lower) since the two corresponding bulk moduli
(orange and red dashed lines in the inset) are very close to each other: in this case the APE is about 1.7\%.
The isothermal QSA bulk modulus (blue dashed line in the inset) is larger so the corresponding TE is smaller
at higher temperatures with an APE $\approx 4.4\%$.
The black line is the TE computed with fixed ECs ($T=0$ K without zero point
contributions).
It gives the smallest TE with an APE of about $7.5\%$.
In general, in silicon the different methods give very similar results, especially at temperature lower than $\approx$
200 K for which an almost perfect overlap is found.
Finally, the theoretical TEs are in good agreement with experimental points~\cite{si_te_1, book3} as already found in
previous literature~\cite{gonze}.

\begin{figure*}
\centering
\includegraphics[width=0.49\linewidth]{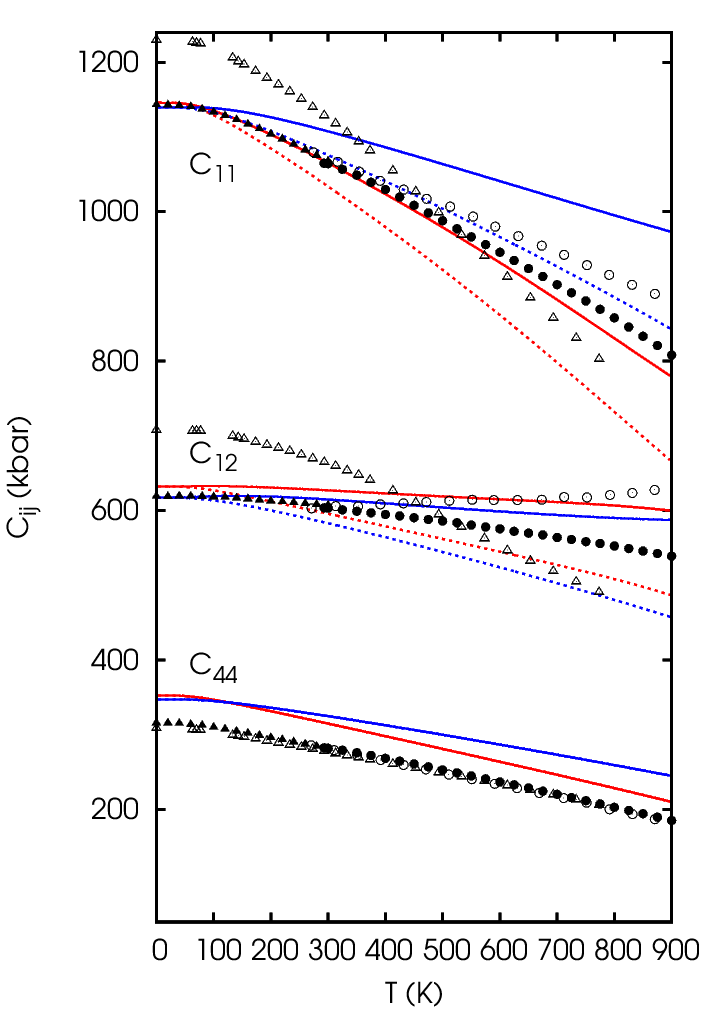}
\caption{Temperature dependent ECs of aluminum. QHA (red curves) is compared with QSA (blue curves). 
        The isothermal (dashed lines) and adiabatic (continuous lines) ECs are reported. Experimental data are taken 
        from: Sutton~\cite{sutton} (open triangles), Kamm and Alers~\cite{kamm} (full triangles), Gerlich and Fisher~\cite{gerlich} 
        (full circles), Tallon and Wolfenden~\cite{tallon} (open circles).}
\label{fig:al_ec}
\end{figure*}

\begin{figure*}
\centering
\includegraphics[width=.49\linewidth]{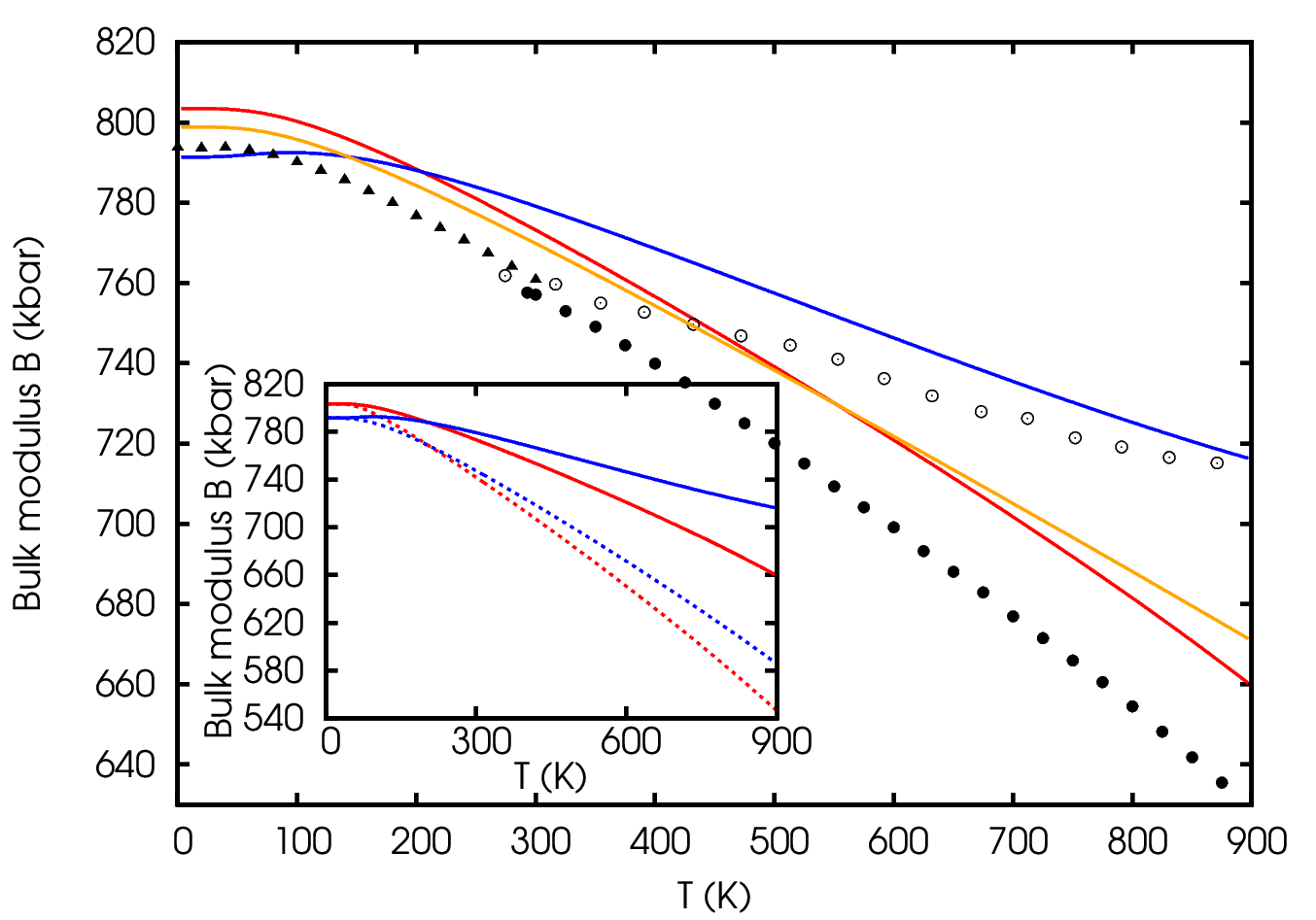}
\includegraphics[width=.49\linewidth]{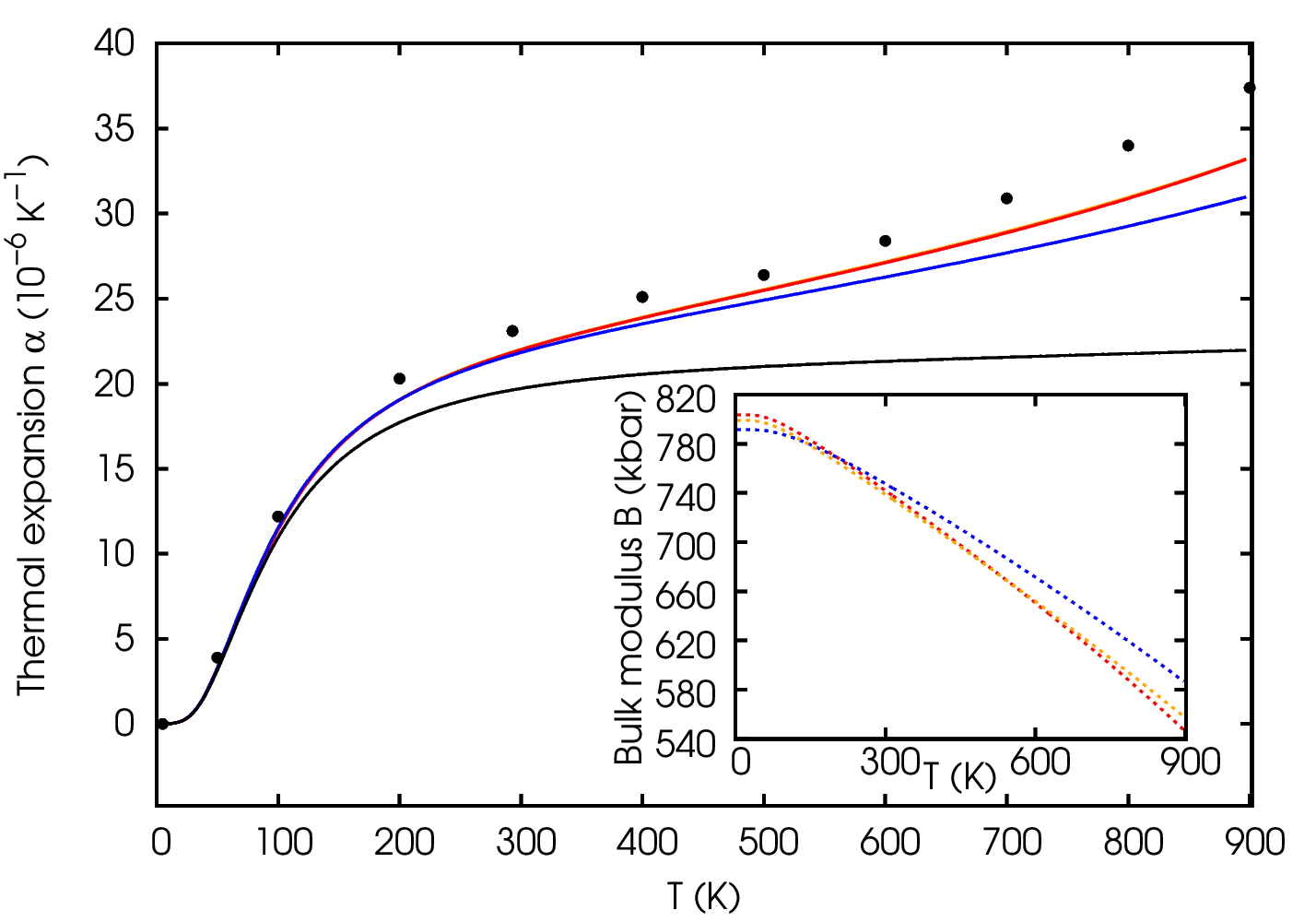}
\caption{Left. Temperature dependent bulk modulus of aluminum. QHA (red curves), QSA (blue curves) and experimental points:
        Kamm and Alers~\cite{kamm} (full triangles), Gerlich and Fisher~\cite{gerlich} (full circles), 
        Tallon and Wolfenden~\cite{tallon} (open circles).
        The orange curve is the bulk modulus obtained from the Murnaghan equation. In the inset a comparison 
        between the adiabatic (continuous lines) and the isothermal(dashed lines) bulk moduli.
        Right. Thermal expansion of aluminum: computed as Eq.~\ref{alfa} (orange curve). The other curves are computed by using 
        Eq.~\ref{alfagrun} with a bulk modulus derived from ECs (Eq.~\ref{bmod}): temperature dependent  
        ECs computed via QHA (red curve) or QSA (blue curve) and $T=0 K$ ECs (black curve). 
        The experimental points are taken from~\cite{book_metal}. In the inset the isothermal bulk moduli
        are reported: QHA (red dashed lines), QSA (blue dashed line) and obtained from Murnaghan equation (orange dashed line).}
\label{fig:al_te}
\end{figure*}

\begin{figure*}
\centering
\includegraphics[width=0.49\linewidth]{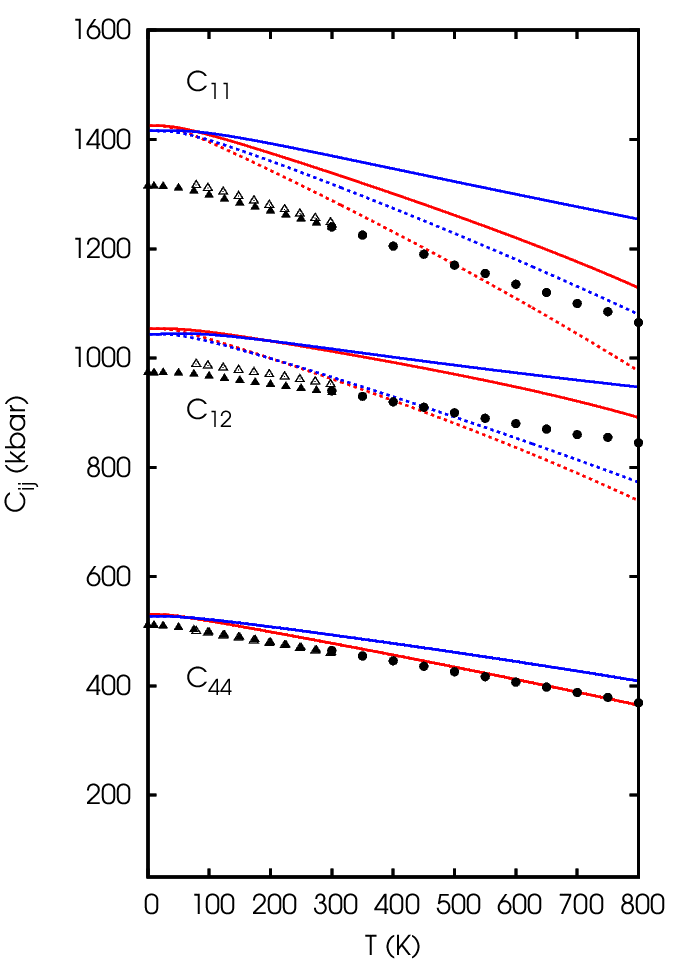}
\caption{Temperature dependent elastic constants of silver. Quasi-harmonic approximation (red curves) is compared with 
        quasi-static approximation (blue curves). The adiabatic (continuous lines) and isothermal (dashed lines) elastic
        constants are reported. Experimental data are taken from Neighbours and Alers~\cite{Neighbours_1958} (full triangles),
        Mohazzabi~\cite{mohazzabi_1984} (open triangles) and Chang and Himmel~\cite{Chang_1966} (circles).}
\label{fig:ag_ec}
\end{figure*}

\begin{figure*}
\centering
\includegraphics[width=.49\linewidth]{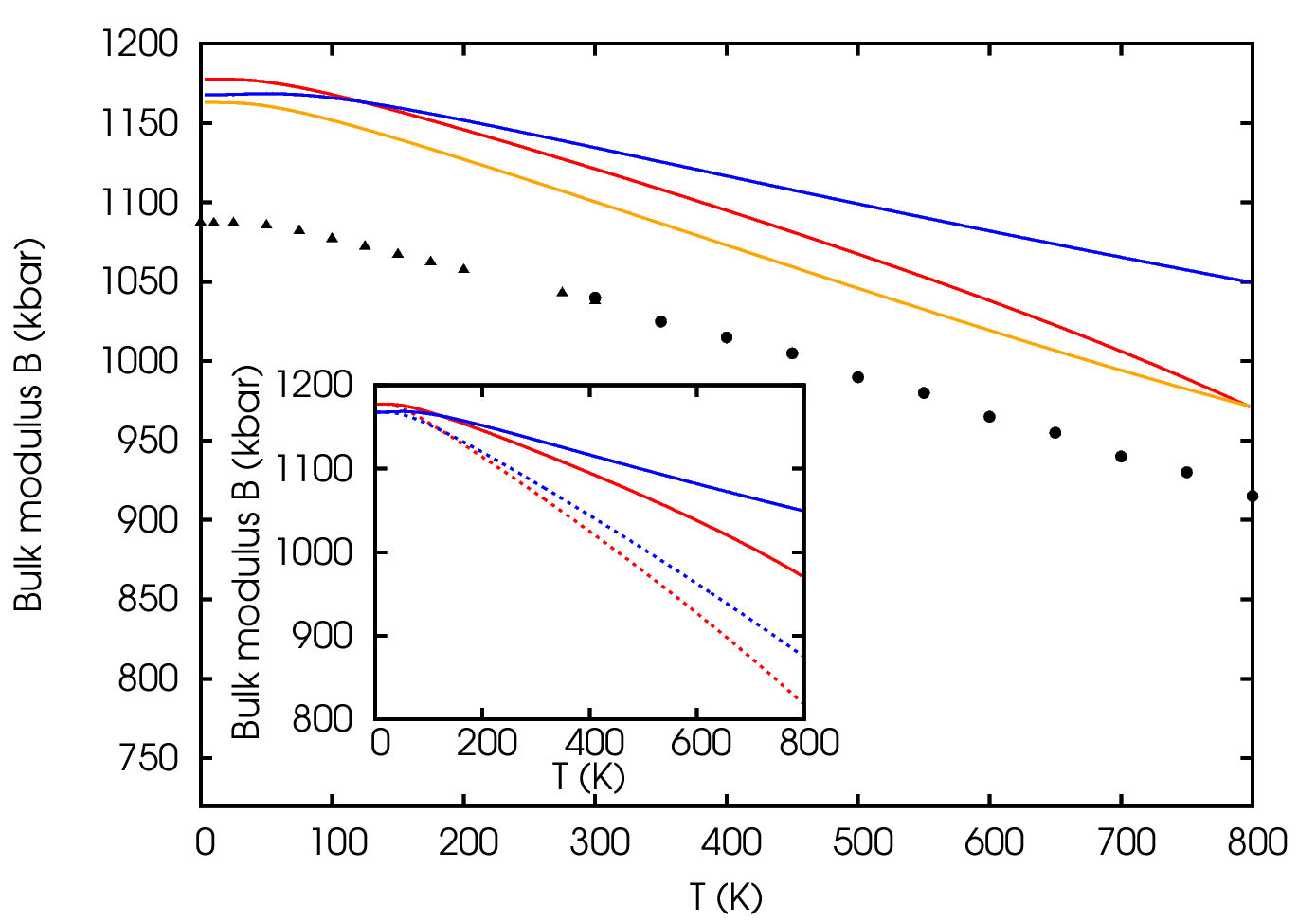}
\includegraphics[width=.49\linewidth]{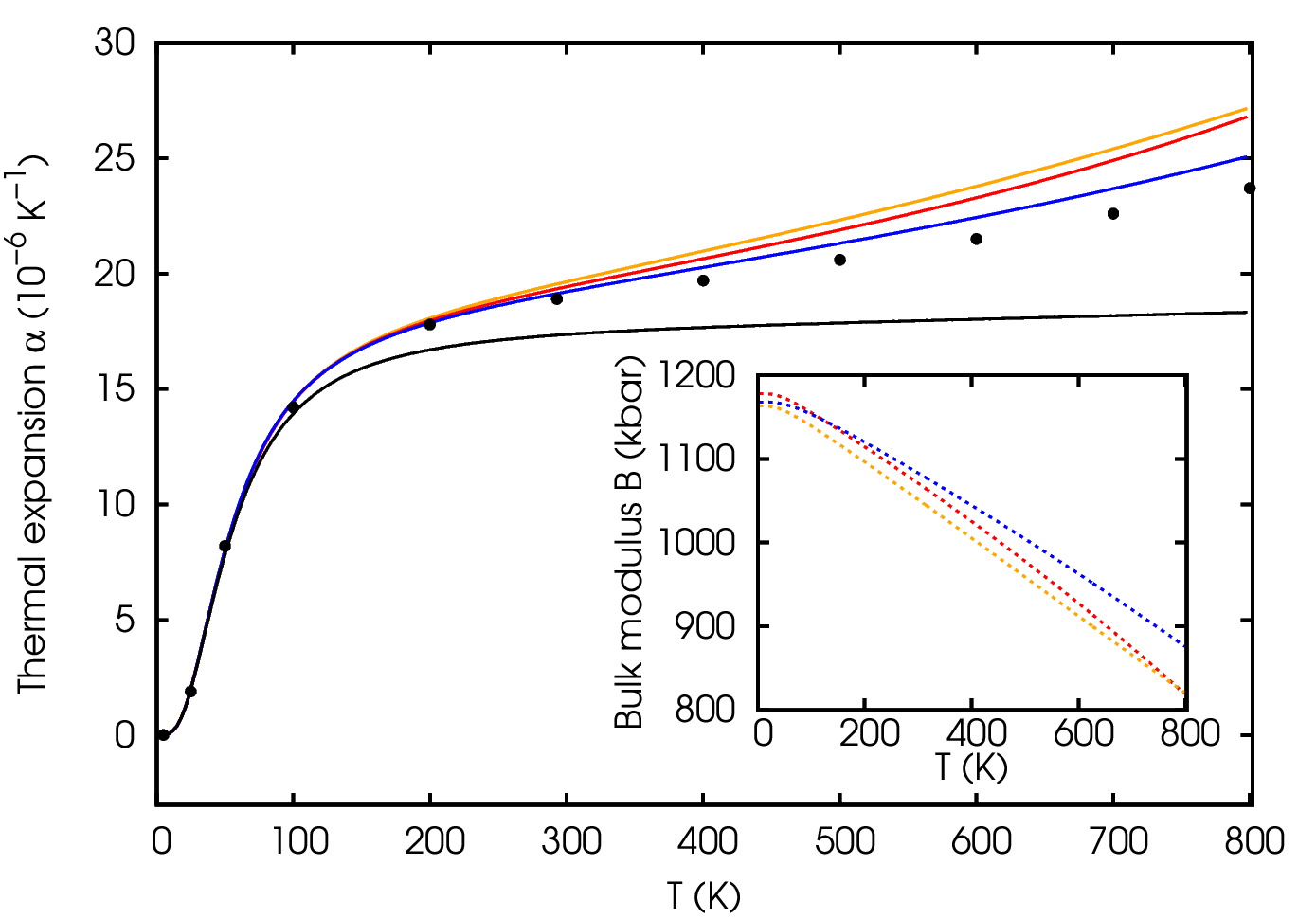}
\caption{Left. Temperature dependent adiabatic bulk modulus of silver. QHA (red curves), QSA (blue curves) and experimental points 
        ~\cite{Neighbours_1958} (full triangles), ~\cite{mohazzabi_1984} (open triangles) and ~\cite{Chang_1966} (circles). 
        The orange curve is the bulk modulus obtained from the Murnaghan equation. In the inset we compare the adiabatic (continuous lines) 
        and the isothermal (dashed lines) bulk moduli.       
        Right. Thermal expansion of silver: computed as Eq.~\ref{alfa} (orange curve). The other curves are computed by using Eq.~\ref{alfagrun} 
        with a bulk modulus derived from ECs (Eq.~\ref{bmod}): temperature dependent ECs computed via QHA (red curve) or 
        QSA (blue curve) and T=0 ECs (black curve). 
        The experimental points are taken from~\cite{book_metal}. In the inset the isothermal bulk moduli are reported: QHA 
        (red dashed line), QSA (blue dashed line) and obtained from Murnaghan equation (orange dashed line).}
\label{fig:ag_te}
\end{figure*}

In Figure~\ref{fig:al_ec} we report TDECs of aluminum with the same meaning
for the lines and colors as for silicon.
In aluminum there are several ultrasonic experiments which do not totally agree
with each other.
We report these experimental points in the same figure.
Sutton data~\cite{sutton} are in the temperature range $63$ K - $773$ K
(open triangles),
Kamm and Alers~\cite{kamm} data are in the range $4.2$ K - $300$ K (full triangles),
Gerlich and Fisher~\cite{gerlich} data are in the range $293$ K - $925$ K
(full circles), and
Tallon and Wolfenden~\cite{tallon} data are in the range
$273$ K - $913$ K (full circles).
All experimental ECs are adiabatic and must be compared with the continuous lines.
The temperature dependence and also the actual values measured by Sutton are quite distant
from the other measurements for $C_{11}$ and $C_{12}$ (while the agreement improves for $C_{44}$).
For this reason we do not further discuss these data in the rest of the paper.
The QHA is in satisfactory agreement with both the data of Kamm and Alers, and Gerlich and Fisher.
The data of Tallon and Wolfenden indicate less softening in $C_{11}$ than Gerlich and Fisher and
report a $C_{12}$ which is approximately constant within the experimental errors, very similar to
the theory in this case.
Moreover, recent measurements of resonant ultrasound spectroscopy~\cite{pham} of Young's modulus
and shear modulus found good agreement with those derived from Gerlich and Fisher.
On the other hand, the QSA shows a smaller softening in $C_{11}$ and $C_{44}$ than the QHA,
while smaller differences between the two approximations are present for $C_{12}$.

The percentage differences between theoretical ECs (red continuous curves) and experimental points at $T=0$ K (Kamm data) are
$-0.2$ \% (QHA) and $0.3$ \% (QSA) for $C_{11}$, $-2$ \% (QHA) and $0.3$ \% (QSA) for $C_{12}$, $-11$ \% (QHA) and $-10$ \%
(QSA) for $C_{44}$. The $T=0$ K ECs are also reported in Table~\ref{table2}
and compared with previous literature.
In the temperature range $0$ K - $800$ K the experiment of Kamm and Alers, and
Gerlich and Fisher taken together show a softening of $\approx 25 \%$ for $C_{11}$,
$\approx 11 \%$ for $C_{12}$ and $\approx 36 \%$ for $C_{44}$
While taking together the experiments of Kamm and Alers, and Tallon and Wolfenden in the same range of temperatures the percentage
variations are: $\approx 20 \%$ for $C_{11}$, $\approx -0.3 \%$ for $C_{12}$ and
$36\ \%$ for $C_{44}$.

The theoretical softening in the same temperature range are $\approx 28 \%$ (QHA) and $\approx 13 \%$
(QSA) for $C_{11}$, $\approx 4 \%$ (QHA and QSA) for $C_{12}$, $\approx 35 \%$ (QHA) and $\approx 25 \%$ (QSA) for $C_{44}$.
As for silicon, in aluminum the QHA reproduces better the experimental trend than QSA.
Similar results have been obtained in Reference~\cite{pham} with the PW functional.

We report in Figure~\ref{fig:al_te} (left) the bulk modulus of aluminum computed with the different sets of ECs with
the same meaning for the lines and colors as for silicon.
The adiabatic bulk modulus derived from the Murnaghan equation is also reported (orange curve):
it is in good agreement with the bulk modulus computed from QHA TDECs
(red curve) while larger differences are present with the QSA (blue curve) which remains higher at
larger temperatures.
In order to check the consistency of our ECs we calculate the TE by using Gr\"{u}neisen's formula using the isothermal bulk modulus obtained from the TDECs (the comparison between adiabatic and isothermal
bulk moduli is shown in the inset). We take as reference the TE coefficient computed from finite differences and compute
the APE as done for silicon.
The TE obtained from mode-Gr\"{u}neisen parameters (Eq.~\ref{alfagrun}) with the bulk modulus derived from the Murnaghan
equation (orange dashed line in the inset) gives an APE $\approx 0.01\%$.
The result of the
TE derived from QHA TDECs (red curve) gives a remarkable agreement with the
reference with an APE $\approx 0.2\%$, while
an APE $\approx 2.4\%$ is found using QSA.
Finally, the calculation of the TE with a fixed bulk modulus obtained
from the $T=0$ KECs has
an APE $\approx 17.2 \%$, even if the agreement remains good at temperature up
to $100$ K.

In Figure~\ref{fig:ag_ec} we report the TDECs of silver with the same meaning for the
lines and colors of previous plots.
The points are taken from ultrasonic experiments that provide adiabatic ECs and must be compared with the continuous lines:
reference~\cite{Neighbours_1958} is in the range of temperatures
$4.2$ K - $300$ K (full triangles), reference ~\cite{mohazzabi_1984} is
in the range $79$ K - $298$ K (open triangles)
and reference~\cite{Chang_1966} is in the range $300$ K - $800$ K (circles). Since the two sets of data below room
temperature are very similar, in the following we limit the comparison with the first one.
The percentage differences between QHA ECs (red continuous curves) and experimental points at $T=0$ K are about $8$ \%
for $C_{11}$ and $C_{12}$ and $4$ \% for $C_{44}$. Almost the same values are
found for QSA ECs.
Our values of the $T=0$ K ECs are between the ECs of Refs.~\cite{Wang_2010}
and Ref.~\cite{Shang_2010} both using GGA functionals and are
slightly closer to experiment (see Table~\ref{table2}).
The experiments show a softening of $\approx 19 \%$ for $C_{11}$, $\approx 13 \%$ for $C_{12}$, and $\approx 28 \%$ for
$C_{44}$ in the temperature range $0$ K - $800$ K. In the same range of temperatures the theoretical softening of the adiabatic ECs
is $\approx 21 \%$ (QHA) and $\approx 11 \%$ (QSA) for $C_{11}$, $\approx 15 \%$ (QHA) and $\approx 9 \%$ (QSA)
for $C_{12}$ and $\approx 31 \%$ (QHA) and $\approx 23 \%$ (QSA) for $C_{44}$.
As for silicon and aluminum the QHA reproduces slightly better the experimental trend.
We report in Figure~\ref{fig:ag_te} (left) the adiabatic bulk modulus of silver computed with the different sets of ECs
and the one derived from the Murnaghan equation (orange curve).
The QHA bulk modulus is the closest to the orange curve, even if
a small artificial descent is present in the high temperature part of the red curve ($>$ 600 K).
This may be due to a not perfect convergence in the {\bf k}-point sampling but since the variation is small and the selected
{\bf k}-point grid was already close to the largest one we could afford we have not investigated further this issue.
Finally we check the consistency of our ECs by computing the TE using the Gr\"{u}neisen's formula and
the isothermal bulk modulus obtained from the TDECs (the comparison between adiabatic and isothermal
bulk moduli is shown in the inset).
The APE is computed as for the previous two materials.
The TE computed
using the isothermal bulk modulus derived from the Murnaghan equation (orange dashed curve in the inset) has an APE of the
order of $-0.01\%$, being practically identical to the finite differences method.
Also QHA TDECs give a very good agreement, the associated APE is $\approx 1.4\%$, while the QSA TE has slightly smaller
values at higher temperatures with an APE of $\approx 4.2\%$.
Finally, by using a temperature independent bulk modulus obtained from the
$T=0$ K ECs (black line) the TE is more distant from the reference
with an APE $\approx 18.4 \%$.

\section{Conclusions}

In this paper we presented our implementation of the TDECs in the \texttt{thermo\_pw} code.
Isothermal and adiabatic ECs can be computed within the framework of DFT and DFPT within
both the QSA and the QHA. The two approaches have been compared in this paper.
The first method is less computationally demanding and takes into account only the
TE effect.
The second method, more accurate, obtains the ECs from the derivatives
of the Helmholtz free-energy with respect to the strain and requires the
computation of the phonon dispersions in all strained geometries.
The two methods were validated computing the TDECs of three cubic
crystals: silicon, aluminum, and silver. The computed ECs were systematically compared
with experiment. It was found that the experimental temperature dependence agrees very
well with QHA and also QSA gives often reasonable results. The actual values of the
ECs depend on the exchange and correlation functionals: with our choices
the values are usually within 10 \% from experiment
but for some element and EC component the error can be even lower (for instance
it is $\approx 0.3$ \% for $C_{11}$ in aluminum below room temperature).

As a further check we use the TDECs to estimate the TE using the mode
Gr\"{u}neisen parameters.
The QHA ECs allow to reproduce with remarkable accuracy the TE given by
differentiation of the temperature dependent lattice constant.

The method is currently available and tested for cubic solids.
As in many theoretical calculations of the TDECs presented so far,
internal degrees of freedom are optimized only at $T=0$ K negleting the
dependence of the free energy on them.
The extension of the method to anisotropic solids and to a complete QHA
treatment of the internal degrees of freedom will be the subject of
future investigations.

\ack 

Computational facilities have been provided by SISSA through its Linux
Cluster and ITCS and by CINECA through the SISSA-CINECA 2018-2020
Agreement.

\section*{References}

\bibliography{unsrt}{}
\bibliographystyle{unsrt}
 
\clearpage

\end{document}


\title[]{Supplementary material for Quasi-harmonic temperature dependent
elastic constants: applications to silicon, aluminum, and silver
}

\author{C. Malica$^1$ and A. Dal Corso$^{1,2}$}

\address{$^1$International School for Advanced Studies (SISSA), \\
Via Bonomea 265, 34136 Trieste (Italy).\\
$^2$IOM-CNR 34136 Trieste (Italy).}
\ead{cmalica@sissa.it}
\vspace{10pt}
\begin{indented}
\item[]February 2020
\end{indented}


%

\section{Gr\"{u}neisen parameters}
\begin{figure}[]
\begin{minipage}{.6\linewidth}
\centering
\subfloat[]{\label{main:a}\includegraphics[scale=.17]{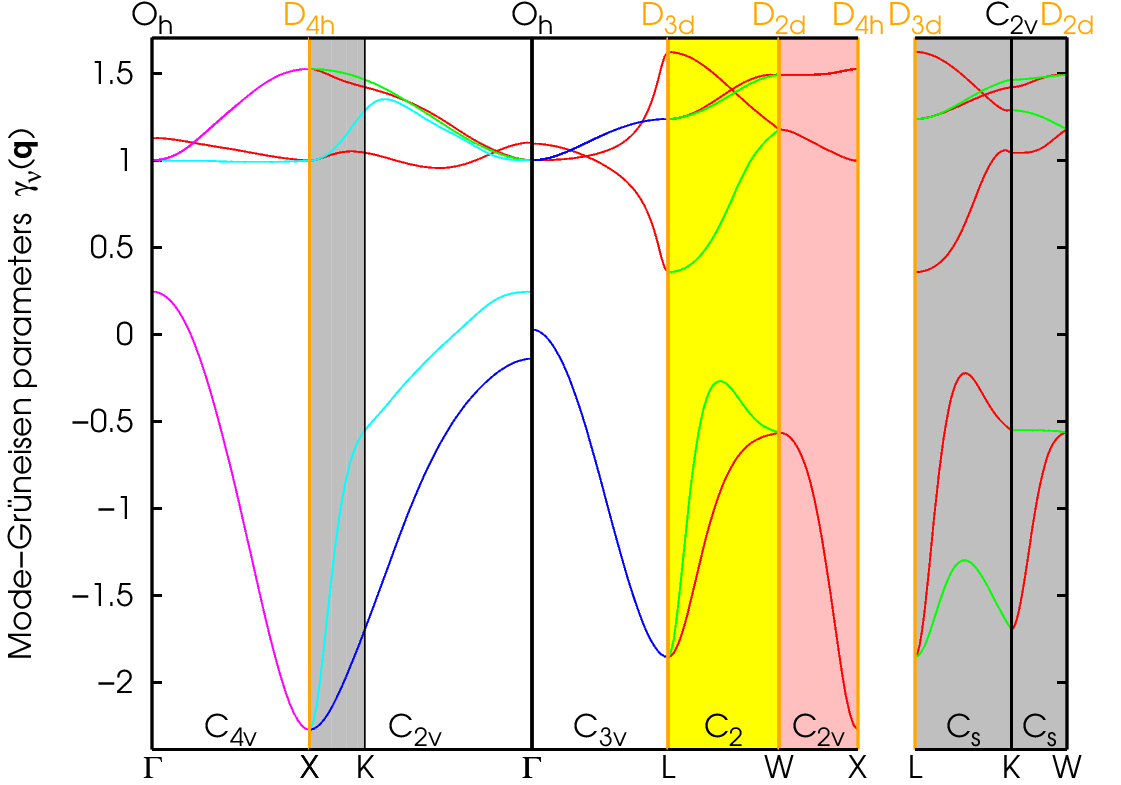}}
\end{minipage}%
\begin{minipage}{.6\linewidth}
\centering
\subfloat[]{\label{main:b}\includegraphics[scale=.17]{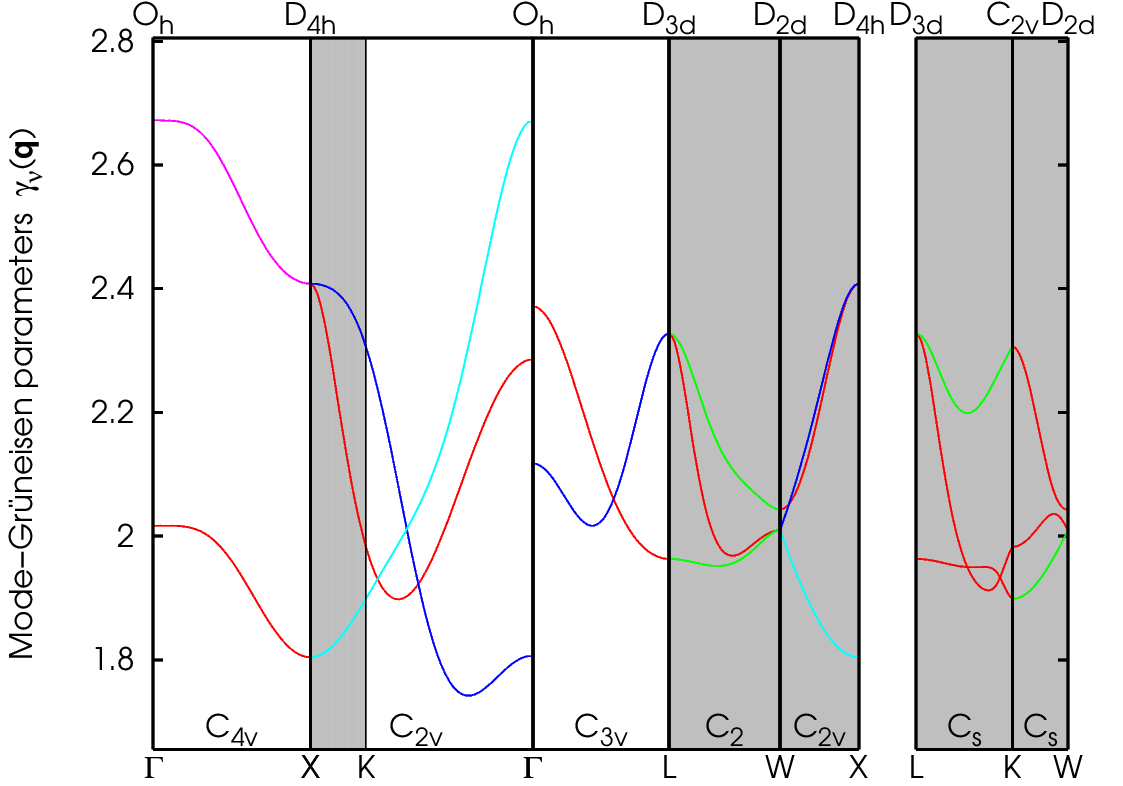}}
\end{minipage}\par\medskip
\centering
\subfloat[]{\label{main:c}\includegraphics[scale=.17]{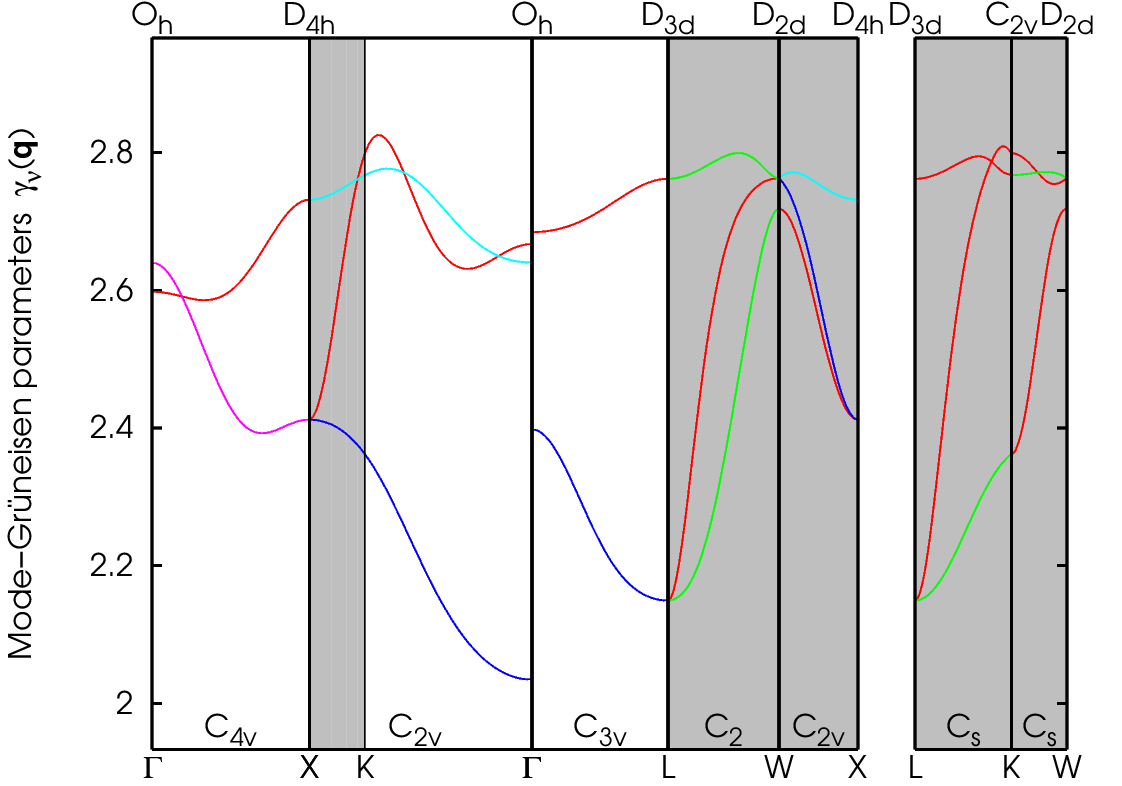}}

\caption{Mode-Gr\"{u}neisen parameters for: a) silicon, b) aluminum, c) silver.}
\label{fig_gp}
\end{figure}

In Figure~\ref{fig_gp} we report the mode-Gr\"{u}neisen parameters
used for the thermal expansion (TE) (Eq. 14) calculation for silicon, aluminum, 
and silver, 
calculated along selected high symmetry lines in the Brillouin zone.

\section{Temperature dependent elastic constant flow-chart}

\begin{figure}
\begin{minipage}{.7\linewidth}
\subfloat[]{\label{main1:a}\includegraphics[scale=.3]{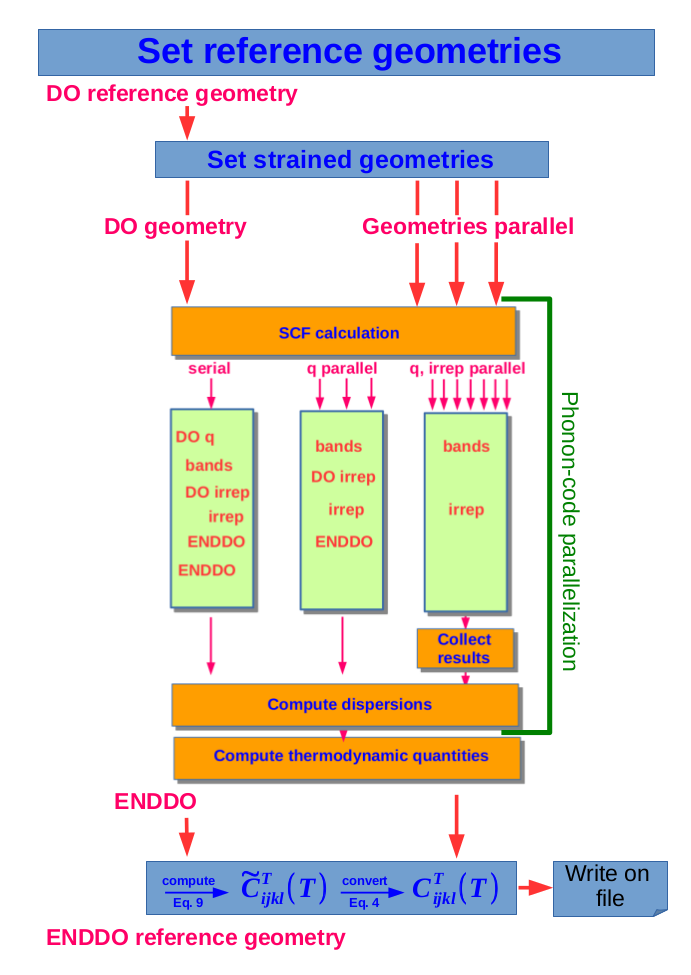}}
\end{minipage}
\begin{minipage}{.7\linewidth}
\subfloat[]{\label{main1:b}\includegraphics[scale=.3]{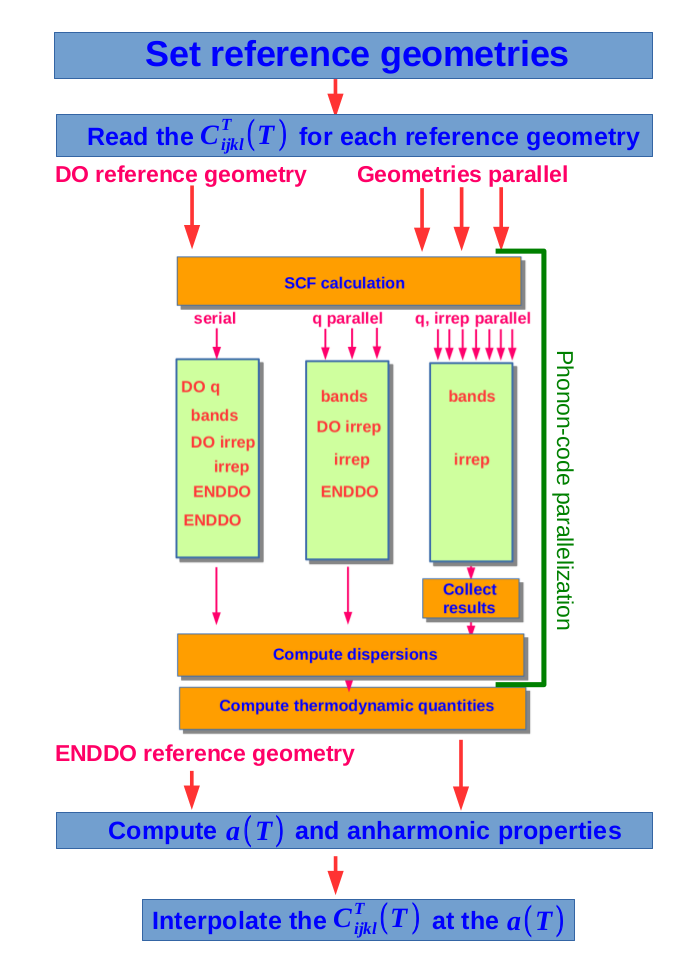}}
\end{minipage}
\caption{Flow-chart of the calculation of the TDECs within the QHA with the \texttt{thermo\_pw} code.}
\label{fig_diag}
\end{figure}

The calculation of the temperature dependent elastic constants (TDECs) 
within the QHA with the \texttt{thermo\_pw} 
code is divided in two parts illustrated in Figure~\ref{fig_diag}.

The first part (a) corresponds to the calculation of the $C^T_{ijkl}$ 
as a function of the temperature for the reference geometries given in input. 
For each reference geometry all the strained geometries required for an elastic constant
calculation are generated and phonons are computed for each strained 
configuration. All these calculations can be ran sequentially or 
in parallel. The parallelism underlying each phonon-dispersion 
calculation (see for instance: ICTP lecture notes {\bf 24}, 
163 (2009) by R. di Meo {\it et al.}) is implemented in {\tt thermo\_pw}
using images parallelization. Having the phonon frequencies of each strained
geometry it is possible to compute, within the harmonic approximation, the 
thermodynamic quantities that depend from them (in particular the vibrational free
energy of Eq. 10). Then the second derivatives of the free-energy with
respect to strain $\tilde C^T_{ijkl}$ are computed at
each temperature (Eq. 9) and the results are corrected for finite pressure 
effects in order to get the stress-strain ECs $C^T_{ijkl}$
(Eq. 4): this in shown in the last block in the figure. 
A file with the $C^T_{ijkl}(T)$ is written for each reference geometry.

The second part (b) is devoted to the computation of the anharmonic properties 
within the quasi-harmonic approximation. It needs the files of the $C^T_{ijkl}(T)$
at each reference geometry produced by the 
first part (a). The setup of the references geometries must be the same as in
part (a).
Phonons are computed at each reference geometry. This allows to compute the 
vibrational free-energy of each reference geometry. 
For each temperature $T$, the vibrational free-energy is added to the energy
to get the Helmholtz free energy of the solid at the reference geometries.
The free-energies at the various geometries are interpolated by
a polynomial (the order of the polynomial can be varied) and minimized to obtain 
$a(T)$. 
The isothermal $C^T_{ijkl}(T)$ at the various geometries are interpolated 
by a polynomial and, at each temperature $T$, the value corresponding to $a(T)$ is 
evaluated from the polynomial interpolation: in this way we obtain the final QHA TDECs.
Then, starting from this isothermal QHA TDECs, the adiabatic QHA TDECs,
elastic compliances and  bulk modulus are derived. In this second part 
the parallellization is done as
in the first part as shown in Figure~\ref{fig_diag} although the number
of required phonon dispersions is much smaller than in 
the first part. 
In order to compute the free-energy as a function of the strain we used in the first part an even number 
(6 values) of strains centered around $\epsilon=0$.
However this point is not included. If the user select an odd number of strained configurations 
then the central geometry is considered. In this case the user can use the dynamical matrices already computed for 
the reference $\epsilon=0$ geometry to run the second part of the calculation.

In order to compute the TDECs within the QSA, the ECs are computed at
$T=0$ K at the
different reference geometries by using the stress-strain relation (Eq. 2) or
the second derivatives of the total energy (Eq. 3 and Eq. 4) and
then they are saved on a different file for each geometry.
Part (b) does not change provided that the files of the $T=0$ K ECs are 
read instead of the $C^T_{ijkl}(T)$.

\section{Some tests: elastic constants at $T=0$ K}

Some tests of the ECs calculation by using the {\tt thermo\_pw} code 
can be found in Ref. [21].  
In this section we investigate a few other examples. 
As in Ref. [4] we considered In, TiO$_2$ rutile, and Al$_2$O$_3$ 
and computed the ECs at $T=0$ K from the second derivatives of
the total energy with respect to strain (Eq. 3 of the paper), after
minimizing the total energy with respect to the crystal parameters and
finding the equilibrium geometry.
We also verified the result computing the ECs from the stress-strain 
relation (Eq. 2 of the paper). Both methods are available in the
code. Moreover, we computed some properties of macroscopic elasticity
as described in the tables captions. In these tables
we compare our results with those obtained in Ref. [4]
and with the experimental data reported in the same paper.

\subsection{Indium}
We used the Wu-Cohen (WC) exchange-correlation functional 
(Z. Wu, R.E. Cohen, Phys. Rev. B {\bf 73}, 235116, 2006) and
the pseudopotential \texttt{In.wc-dn-}\texttt{kjpaw\_psl.1.0.0.UPF}
from {\tt pslibrary}. The cutoff for the wave functions
was $70$ Ry, the one for the charge density $500$ Ry, 
the \textbf{k}-point mesh was $48 \times 48 \times 32$.
The presence of the Fermi surface has been dealt with by the MP [41]
smearing technique with a value of the smearing parameter $\sigma=0.02$.
The equilibrium configuration was obtained by interpolating the total
energy computed in a $5\times5$ grid of $a$ and $c/a$ crystal parameters
with a 2-dimensional fourth-degree polynomial and by minimizing it
($\Delta a = 0.05$ a.u., $\Delta (c/a) = 0.02$). The results are reported in 
Table~\ref{table1}.

\begin{table*}\centering
\caption{
Elastic properties of indium. Crystal parameters are
in units of the Bohr radius. The elastic constants $C_{ij}$ are in kbar.
Bulk modulus B (kbar), shear modulus S (kbar) and Young's modulus Y (kbar) and
Poisson's ratio V are calculated within the Voigt (V) and Reuss (R) 
approximations. The average of the two according to the 
Voigt-Reuss-Hill (H) method is also reported.
Transverse elastic wave velocity $v_t$,
longitudinal elastic wave velocity $v_l$ and the
average wave velocity $v_m$ are reported in m/s and the Debye temperature
$\Theta_D$ in K.
}
\begin{tabular}{lccc}
\hline
& \multicolumn{1}{c}{This work} \ \ & \multicolumn{1}{c}{Other theoretical} \ \ 
& \multicolumn{1}{c}{Expt.} \\
\hline
\hline
Method \ & PAW-PP & FP-LAPW & $$ \\
Functional \ & WC & WC & $$ \\
$a_0$ \ & $6.0956$ & $6.0491$ & $6.1439$ \\
$c_0$ \ & $9.3414$ & $9.4324$ & $9.3479$ \\
$C_{11}$ \ & $617$ & $589$ & $525$ \\
$C_{12}$ \ & $302$ & $332$ & $368$ \\
$C_{13}$ \ & $424$ & $374$ & $371$ \\
$C_{33}$ \ & $500$ & $448$ & $530$ \\
$C_{44}$ \ & $70$ & $58$ & $78$ \\
$C_{66}$ \ & $54$ & $25$ & $147$ \\
$B_{V}$ \ & $448$ & $421$ & $422$ \\
$B_{R}$ \ & $448$ & $414$ & $422$ \\
$B_{H}$ \ & $448$ & $418$ & $422$ \\
$S_{V}$ \ & $78$ & $65$ & $92$ \\
$S_{R}$ \ & $62$ & $50$ & $86$ \\
$S_{H}$ \ & $70$ & $57$ & $89$ \\
$Y_{V}$ \ & $220$ & $185$ & $257$ \\
$Y_{R}$ \ & $178$ & $143$ & $242$ \\
$Y_{H}$ \ & $199$ & $164$ & $250$ \\
$V_{V}$ \ & $0.418$ & $0.426$ & $0.398$ \\
$V_{R}$ \ & $0.434$ & $0.442$ & $0.404$ \\
$V_{H}$ \ & $0.425$ & $0.434$ & $0.401$ \\
$v_{t}$ \ & $971$ & $875.2$ & $1105.2$ \\
$v_{l}$ \ & $2702$ & $2573.2$ & $2723.6$ \\
$v_{m}$ \ & $1103.1$ & $995.4$ & $1251.4$ \\
$\Theta_{D}$ \ & $109$ & $100.6$ & $125.5$ \\
\hline
\hline
\end{tabular}
\label{table1}
\end{table*}

\subsection{Rutile $Ti O_2$}
We used the Perdew-Burke-Ernzerhof (PBE) exchange-correlation functional 
(J. P. Perdew {\it et al.},, Phys. Rev. Lett. {\bf 77}, 3865, 1996)   and
the pseudopotentials \texttt{Ti.pbe-spn-}\texttt{kjpaw\_psl.1.0.0.UPF} and
\texttt{O.pbe-nl-}\texttt{kjpaw\_psl.1.0.0.UPF} from {\tt pslibrary}. 
The cutoff for the wave functions
was $50$ Ry, the one for the charge density $350$ Ry,
the \textbf{k}-point mesh was $12 \times 12 \times 20$.
The equilibrium configuration was obtained by interpolating the total
energy computed in a $5\times5$ grid of $a$ and $c/a$ crystal parameters
with a 2-dimensional fourth-degree polynomial and by minimizing it
($\Delta a = 0.05$ a.u., $\Delta (c/a) = 0.02$). The results are reported in
Table~\ref{table2}.

\begin{table*}\centering
\caption{
Elastic properties of rutile TiO$_2$. Crystal parameters are
in units of the Bohr radius. The elastic constants $C_{ij}$ are in kbar.
Bulk modulus B (kbar), shear modulus S (kbar) and Young's modulus Y (kbar) and
Poisson's ratio V are calculated within the Voigt (V) and Reuss (R)
approximations. The average of the two according to the
Voigt-Reuss-Hill (H) method is also reported.
Transverse elastic wave velocity $v_t$,
longitudinal elastic wave velocity $v_l$ and the
average wave velocity $v_m$ are reported in m/s and the Debye temperature
$\Theta_D$ in K.
}
\begin{tabular}{lccc}
\hline
& \multicolumn{1}{c}{This work} \ \ & \multicolumn{1}{c}{Other theoretical} \ \
& \multicolumn{1}{c}{Expt.} \\
\hline
\hline
Method \ & PAW-PP & FP-LAPW & $$ \\
Functional \ & PBE & PBE & $$ \\
$a_0$ \ & $8.7860$ & $8.6809$ & $8.6806$ \\
$c_0$ \ & $5.6138$ & $5.5900$ & $5.5911$ \\
$C_{11}$ \ & $2566$ & $2683$ & $2690$ \\
$C_{12}$ \ & $1677$ & $1802$ & $1770$ \\
$C_{13}$ \ & $1465$ & $1464$ & $1460$ \\
$C_{33}$ \ & $4693$ & $4779$ & $4800$ \\
$C_{44}$ \ & $1148$ & $1223$ & $1240$ \\
$C_{66}$ \ & $2111$ & $2236$ & $1920$ \\
$B_{V}$ \ & $2116$ & $2178$ & $2173$ \\
$B_{R}$ \ & $2012$ & $2094$ & $2086$ \\
$B_{H}$ \ & $2064$ & $2136$ & $2130$ \\
$S_{V}$ \ & $1229$ & $1298$ & $1246$ \\
$S_{R}$ \ & $947$ & $978$ & $986$ \\
$S_{H}$ \ & $1088$ & $1138$ & $1116$ \\
$Y_{V}$ \ & $3090$ & $3248$ & $3138$ \\
$Y_{R}$ \ & $2456$ & $2538$ & $2556$ \\
$Y_{H}$ \ & $2773$ & $2898$ & $2851$ \\
$V_{V}$ \ & $0.257$ & $0.251$ & $0.259$ \\
$V_{R}$ \ & $0.297$ & $0.298$ & $0.295$ \\
$V_{H}$ \ & $0.274$ & $0.273$ & $0.276$ \\
$v_{t}$ \ & $5133.0$ & $5174.2$ & $5125.8$ \\
$v_{l}$ \ & $9224.5$ & $9272.1$ & $9228.0$ \\
$v_{m}$ \ & $5716.2$ & $5760.8$ & $5709.0$ \\
$\Theta_{D}$ \ & $754.7$ & $785.6$ & $778.5$ \\
\hline
\hline
\end{tabular}
\label{table2}
\end{table*}

\subsection{Rhombohedral $Al_2 O_3$}
We used the Perdew-Burke-Ernzerhof (PBE) exchange-correlation functional and
the pseudopotentials \texttt{Al.pbe-nl-}\texttt{kjpaw\_psl.1.0.0.UPF} and
\texttt{O.pbe-n-}\texttt{kjpaw\_psl.1.0.0.UPF} from {\tt pslibrary}. 
The cutoff for the wave functions
was $70$ Ry, the one for the charge density $500$ Ry,
the \textbf{k}-point mesh was $12 \times 12 \times 12$.
The equilibrium configuration was obtained by interpolating the total
energy computed in a $5\times5$ grid of the lattice constant $a$ and angle 
$\alpha$ with a 2-dimensional fourth-degree polynomial and by minimizing it
($\Delta a = 0.05$ a.u., $\Delta \alpha = 0.5$). The results are reported in
Table~\ref{table3}.

\begin{table*}\centering
\caption{Elastic properties of rhombohedral Al$_2$O$_3$. Crystal parameters are
in units of the Bohr radius. The elastic constants $C_{ij}$ are in kbar. 
Bulk modulus B (kbar), shear modulus S (kbar) and Young's modulus Y (kbar) and
Poisson's ratio V are calculated within the Voigt (V) and Reuss (R)
approximations. The average of the two according to the
Voigt-Reuss-Hill (H) method is also reported.
Transvers elastic wave velocity $v_t$,
longitudinal elastic wave velocity $v_l$ and the
average wave velocity $v_m$ are reported in m/s and the Debye temperature 
$\Theta_D$ in K.
}
\begin{tabular}{lccc}
\hline
& \multicolumn{1}{c}{This work} \ \ & \multicolumn{1}{c}{Other theoretical} \ \
& \multicolumn{1}{c}{Expt.} \\
\hline
\hline
Method \ & PAW-PP & FP-LAPW & $$ \\
Functional \ & PBE & PBE & $$ \\
$a_0$ \ & $9.0924$ & $9.0928$ & $8.9916$ \\
$c_0$ \ & $24.8081$ & $24.8253$ & $24.5498$ \\
$C_{11}$ \ & $4516$ & $4656$ & $4974$ \\
$C_{12}$ \ & $1510$ & $1383$ & $1640$ \\
$C_{13}$ \ & $1091$ & $1036$ & $1122$ \\
$C_{33}$ \ & $4537$ & $4588$ & $4991$ \\
$C_{44}$ \ & $1319$ & $1363$ & $1474$ \\
$C_{14}$ \ & $-200$ & $-24$ & $-236$ \\
$B_{V}$ \ & $2328$ & $2312$ & $2523$ \\
$B_{R}$ \ & $2325$ & $2308$ & $2518$ \\
$B_{H}$ \ & $2326$ & $2310$ & $2521$ \\
$S_{V}$ \ & $1487$ & $1569$ & $1660$ \\
$S_{R}$ \ & $1443$ & $1547$ & $1606$ \\
$S_{H}$ \ & $1465$ & $1558$ & $1633$ \\
$Y_{V}$ \ & $3677$ & $3838$ & $4084$ \\
$Y_{R}$ \ & $3587$ & $3793$ & $3974$ \\
$Y_{H}$ \ & $3632$ & $3816$ & $4029$ \\
$V_{V}$ \ & $0.237$ & $0.223$ & $0.230$ \\
$V_{R}$ \ & $0.242$ & $0.226$ & $0.236$ \\
$V_{H}$ \ & $0.240$ & $0.224$ & $0.233$ \\
$v_{t}$ \ & $6160.5$ & $6355.4$ & $6399.3$ \\
$v_{l}$ \ & $10529.9$ & $10665.7$ & $10853.7$ \\
$v_{m}$ \ & $6831.3$ & $7035.3$ & $7090.9$ \\
$\Theta_{D}$ \ & $979.4$ & $1015.3$ & $1034.8$ \\
\hline
\hline
\end{tabular}
\label{table3}
\end{table*}

\clearpage

\section{Thermoelasticity of MgO}
In this section we compare the results obtained by our approach with the QHA 
TDECs of MgO available in the literature in Refs. [12, 13] cited in the paper. 
This example is also useful to test a material with ionic bonds,
since in the main paper we considered only covalent and metallic systems. 

\begin{figure}[b]
\centering
\includegraphics[width=0.49\linewidth]{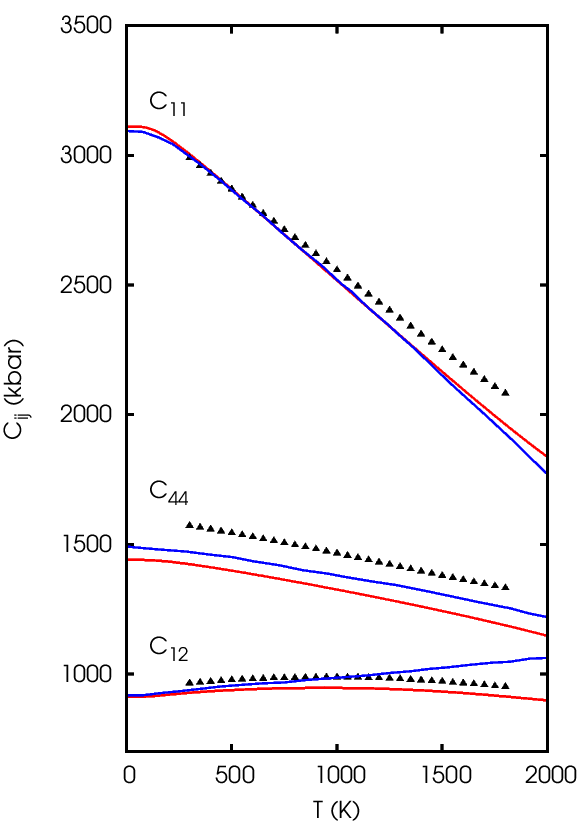}
\caption{Adiabatic QHA TDECs of MgO. This work (red curves) compared with the results of Refs. [12,13] (blue curve). 
        The points are experimental data, the same shown in [12,13]. }
\label{fig:mgo}
\end{figure}

In order to minimize as much as possible the differences between our 
calculation and the literature, we used the LDA for the exchange-correlation 
energy and the norm-conserving pseudopotentials from the 
Quantum Espresso pseudopotentials library: \texttt{Mg.pz-n-vbc.UPF}
and \texttt{O.pz-mt.UPF}. The cutoff for the wave functions was $90$ Ry. 
The \textbf{k}-point mesh was $12 \times 12 \times 12$.
Density functional perturbation theory (DFPT) was used to calculate the dynamical 
matrices on a $4 \times 4 \times 4$ \textbf{q}-point grid.
These dynamical matrices have been Fourier interpolated on a $200 \times 200 \times 200$ 
\textbf{q}-point mesh to evaluate the free-energy.
We used $9$ reference geometries with lattice constants separated from each other by 
$\Delta a= 0.05$ a.u. and centered in the $T=0$ K equilibrium lattice constant: 
$7.93$ a.u.. In order to fit the free-energy as a function of strain we use a 
polynomial of degree two.
To fit the ECs computed at the various reference configurations at the temperature dependent 
geometry we use a polynomial of degree three. In Fig.~\ref{fig:mgo} we report 
the TDECs.

